\pdfoutput=1
\documentclass[aps,prd,twocolumn,groupedaddress,showpacs,nofootinbib]{revtex4-1}

\usepackage{latexsym} 
\usepackage{amssymb} 
\usepackage{amsfonts}
\usepackage{bm}
\usepackage{hyperref}

\usepackage[utf8x]{inputenc}
\usepackage{graphicx} 
\usepackage[squaren]{SIunits} 

\graphicspath{ {./plots/} }

\newcommand{\Gauss}{\ensuremath{\mathrm{G}}}

\begin{document}


\title{Binary Neutron Star Mergers and Short Gamma-Ray Bursts:\\ Effects
  of Magnetic Field Orientation, Equation of State, and Mass Ratio}



\author{Takumu Kawamura}
\author{Bruno Giacomazzo}
\email[]{bruno.giacomazzo@unitn.it}
\author{Wolfgang Kastaun}
\author{Riccardo Ciolfi}
\author{Andrea Endrizzi}
\affiliation{Physics Department, University of Trento, via Sommarive
  14, I-38123 Trento, Italy}
\affiliation{INFN-TIFPA, Trento Institute for Fundamental Physics
  and Applications, via Sommarive 14, I-38123 Trento, Italy}

\author{Luca Baiotti}
\affiliation{Graduate School of Science, Osaka University, Toyonaka,
  560-0043, Japan}

\author{Rosalba Perna} 
\affiliation{Department of Physics and Astronomy, Stony Brook
  University, Stony Brook, New York 11794-3800, USA}


\date{\today}

\begin{abstract}
We present fully general-relativistic magnetohydrodynamic
simulations of the merger of binary neutron star (BNS) systems. We
consider BNSs producing a hypermassive neutron star (HMNS) that
collapses to a spinning black hole (BH) surrounded by a magnetized
accretion disk in a few tens of ms. We investigate whether such
systems may launch relativistic jets and hence power short gamma-ray
bursts. We study the effects of different equations of state (EOSs),
different mass ratios, and different magnetic field orientations. For
all cases, we present a detailed investigation of the matter dynamics
and of the magnetic field evolution, with particular attention to its
global structure and possible emission of relativistic jets.

The main result of this work is that we observe the formation of an
organized magnetic field structure. This happens independently of EOS,
mass ratio, and initial magnetic field orientation. We also show that
those models that produce a longer-lived HMNS lead to a stronger
magnetic field before collapse to a BH. Such larger fields make it
possible, for at least one of our models, to resolve the magnetorotational instability and hence
further amplify the magnetic field in the disk. However, by the end of
our simulations, we do not (yet) observe a magnetically dominated
funnel nor a relativistic outflow. With
respect to the recent simulations of Ruiz et al. [Astrophys. J. 824, L6 (2016)], we
evolve models with lower and more plausible initial magnetic field
strengths and (for computational reasons) we do not evolve the
accretion disk for the long time scales that seem to be required in
order to see a relativistic outflow. Since all our
models produce a similar ordered magnetic field structure aligned with
the BH spin axis, we expect that the results found by Ruiz et al. (who
only considered an equal-mass system with an ideal fluid
EOS) should be general and, at least from a qualitative point of view,
independent of the mass ratio, magnetic field orientation, and EOS.
\end{abstract}



\maketitle


\section{Introduction \label{sec_intro}}

With the revolutionary first detections of gravitational waves (GWs)
by Advanced LIGO~\cite{LIGO:BBHGW:2016, LIGO_GW151226} from the merger
of compact binary systems composed of two black holes (BHs), there
have been even greater expectations of possible near-future detections of
other sources, including binaries composed either of two neutron stars
(NSs) or of a NS and a BH. While solar-mass binary BH mergers are not
expected to emit electromagnetic (EM) signals (but see,
e.g., Refs.~\cite{Perna2016, Janiuk2016, Murase2016} for possible
alternatives), binary neutron star (BNS) and NS-BH systems are
considered very powerful sources of a variety of EM counterparts,
ranging from collimated emission, such as short gamma-ray bursts
(SGRBs), to more isotropic ones, such as the so-called
kilonova/macronova~\cite{Li:1998:L59, Kulkarni:2005:macronova-term,
  Metzger2012}.

In particular, the possibility that SGRBs are powered by BNS or
NS-BH mergers is supported by observational evidence
(see Ref.~\cite{Berger2014} for a recent review). The simultaneous detection
of a SGRB and GWs from a BNS or a NS-BH merger would represent
definitive proof that these binary mergers power the central engine of
SGRBs. Moreover, this association could provide strong constraints 
on the equation of state (EOS) of NS matter~\cite{Giacomazzo2013}.

One of the leading theoretical models describing the gamma-ray
emission in SGRBs is based on the launch of a relativistic jet from a
spinning BH surrounded by an accretion disk. Jets may be launched via
neutrino-antineutrino annihilation~\cite{Narayan1992, Piran:2004:76,
  Nakar:2007:442} or via magnetic mechanisms, such as the
Blandford-Znajek (BZ) mechanism~\cite{Blandford1977}. While fully general
relativistic simulations of BNS mergers have shown that, in those
cases where the merger results in BH formation on a dynamical
time scale, disks as massive as $\sim 0.1 M_{\odot}$ can be easily
formed~\cite{Rezzolla:2010:114105}, whether the emission of
relativistic jets occurs or not is still under investigation.

This has driven an increasing effort in performing fully general-relativistic
magnetohydrodynamic (GRMHD) simulations of BNS mergers, with the first
simulations dating back to a few years
ago~\cite{Anderson2008PhRvL.100s1101A, Liu2008, Giacomazzo2009}. More
recently some groups started to investigate the formation of
jets~\cite{Rezzolla:2011:6, Kiuchi:2014:41502, Ruiz2016, Dionysopoulou:2015:92}. The
simulation by Rezzolla et al.~\cite{Rezzolla:2011:6} was in particular
the first to show the possibility of forming an ordered and mainly
poloidal magnetic field configuration aligned with the BH spin
axis. Even if no outflow was observed, this provided a strong
indication that BNS mergers can at least provide some of the necessary
conditions to launch a relativistic jet. A subsequent simulation by
Kiuchi et al.~\cite{Kiuchi:2014:41502}, using a different EOS, has
challenged that result. Meanwhile, both local and global simulations of
magnetic field evolution in the merger of BNS systems have shown that
very large fields of up to $\sim 10^{16}$ G can be formed during
merger~\cite{Zrake2013, Giacomazzo:2015,
  Kiuchi:2015:1509.09205}. Since it was shown that the formation of a
magnetically dominated region in the BH ergosphere is a necessary
condition for the activation of the BZ
mechanism~\cite{Komissarov2009}, these new results encouraged
further studies. Very recently, GRMHD simulations by Ruiz et
al.~\cite{Ruiz2016} have shown that, when starting with very large
magnetic fields, it is possible to observe the formation of a mildly
relativistic outflow few tens of ms after BH formation. Even if the
initial magnetic fields were unrealistically large, i.e., $\sim
10^{15}$ G, such fields should be produced after merger and therefore
these simulations provide a proof of concept that jets may indeed be
launched. Moreover, these recent simulations have shown that
jets may be launched even when considering magnetic fields confined
inside the NSs.

All previous simulations considered only equal-mass systems and only
two EOSs: ideal fluid~\cite{Rezzolla:2011:6, Ruiz2016} or piecewise
polytropic~\cite{Kiuchi:2014:41502}. In this paper we extend the
previous investigations by studying, with our GRMHD code
Whisky~\cite{Giacomazzo:2007:235, Giacomazzo2011PhRvD..83d4014G,
  Giacomazzo2013ApJ...771L..26G}, the magnetic field structure that is
formed after the merger of BNS systems and how it depends on the
initial mass ratio, EOS, and initial magnetic field orientation. As
such, our work allows us to assess the robustness of previous results
when these important parameters are changed and we consider this as a
preliminary step before performing simulations with very high
resolutions or using our subgrid model~\cite{Giacomazzo:2015} to
further study the effect of large magnetic field amplifications. All
our simulations start with plausible values for the initial magnetic
field, i.e., $\sim 10^{12}$ G. The role of neutrino emission is not
included in our simulations and we believe that this does not affect
our results qualitatively. We are currently working on the
implementation of neutrino treatment in our GRMHD code and we point
out that up to now only one recent work has presented GRMHD
simulations of BNS merger including magnetic fields, neutrino
emission, and a finite-temperature EOS~\cite{Palenzuela2015}.

Our paper is organized as follows. In Sec.~\ref{sec_numerical_methods}
we describe our numerical setup and in Sec.~\ref{sec_initial_data} we describe the
initial data used in our simulations. We remark that our equal-mass
models are the same as those that were evolved by Rezzolla et
al~\cite{Rezzolla:2011:6} and Kiuchi et
al~\cite{Kiuchi:2014:41502}, while the unequal-mass ones are studied
here for the first time. In Sec.~\ref{evolution} we describe in detail
the evolution of our different initial models, for the first time with
a very accurate description of the magnetic field configurations
formed after merger (implementing also advanced visualization tools
that are described in the Appendix). In Sec.~\ref{sec_SGRB} we discuss
the connection with SGRBs and other possible EM counterparts, while in
Sec.~\ref{sec_GWs} we present the GW signal. In
Sec.~\ref{sec_conclusions} we conclude and summarize the main
results of our work.

We use a system of units in which $G=c=M_{\odot}=1$ unless specified
otherwise. The time is shifted so that $t=0$ refers to the time
of merger, which corresponds to the maximum amplitude in the GW
signal.

\section{Numerical Methods \label{sec_numerical_methods}}

All the simulations discussed in this paper made use of the publicly
available Einstein Toolkit~\cite{Loeffler:2012:115001} coupled with
our fully GRMHD code {\tt Whisky}~\cite{Giacomazzo:2007:235,
  Giacomazzo2011PhRvD..83d4014G, Giacomazzo2013ApJ...771L..26G}.

Our version of the {\tt Whisky} code solves the GRMHD equations on a
dynamically curved background by using the ``Valencia''
formulation~\cite{Anton:2005gi}. In order to satisfy at all times the
divergence-free condition of the magnetic field, we evolve the vector
potential and then recompute the magnetic field from it at each time
step. In order to avoid spurious magnetic field amplifications at the
boundary between refinement levels we use the ``modified Lorenz
gauge'' as described in Refs.~\cite{Etienne:2011re,Farris2012}. The fluxes at
the interfaces between numerical cells are computed using the HLLE
approximate Riemann solver~\cite{Harten:1983:35} that takes as input
the values of the primitive variables reconstructed with the
piecewise-parabolic method~\cite{Colella:1984:174}. We also set the
floor value for the rest-mass density $\rho$ to $10^{-13} \approx
6.2 \times 10^{4} \mathrm{g\, cm^{-3}}$. When $\rho$ decreases below
that level, we reset it to the floor value (which we also call the
artificial atmosphere) and we also set the velocity to be zero. After
BH formation we excise the hydrodynamic variables in the region inside
the apparent horizon (by setting them to the values they have in the artificial
atmosphere) in order to prevent failures in the conservative to
primitive routines due to the high-level of magnetization that may be
reached inside the BH. More technical details about our GRMHD {\tt
Whisky} code can be found in our previous
publications~\cite{Giacomazzo:2007:235, Giacomazzo2011PhRvD..83d4014G,
Giacomazzo2013ApJ...771L..26G, Endrizzi2016}.

In this work, the {\tt Whisky} code is coupled with version {\tt ET\_2014\_05}
(codename ``Wheeler'') of the Einstein Toolkit. The latter
is a collection of publicly available routines for numerical
relativity simulations on supercomputers. In particular, for the
evolution of the spacetime we used the
BSSNOK~\cite{Baumgarte:1998:24007,Shibata:1995:5428,Nakamura:1987:1}
formulation as implemented in the McLachlan code. We also used the
adaptive mesh refinement driver Carpet with a total of six
refinement levels. The finest grids cover each of the NSs during the
inspiral and, after merger, they are merged into a larger one that
covers the resulting hypermassive NS (HMNS). We adopted a resolution on the finest
grids of $\approx 222$ m in the runs using an ideal-fluid EOS and of $\approx
186$ m in the runs using the H4 EOS. This choice has been made so that
the NSs are covered by approximately the same number of points in both
cases. The external boundary is located at a distance of $\approx 1400$ km
in the ideal-fluid case and $\approx 1200$ km in the H4 case. All the
simulations employed reflection symmetry across the equatorial
plane to reduce computational costs.

\begin{table}
  \caption{Initial data parameters: mass ratio ($q=M_\mathrm{g}^1/M_\mathrm{g}^2$),
    total baryonic mass of the system ($M_\mathrm{b}^\mathrm{tot}$), baryonic and 
    gravitational masses of each star at infinite separation 
    ($M_\mathrm{b}$ and $M_\mathrm{g}$), compactness ($M_\mathrm{g}/R_c$,
    dimensionless), initial orbital frequency
    and proper separation ($f_0$ and $d$), initial
    magnetic energy ($E_{B}$), initial maximum value of
    magnetic field strength ($B_\mathrm{max}$), and $A_b$, 
    the value in geometric units used in equation~\ref{eq:Avec} 
    in order to fix $B_\mathrm{max}$.}

  \begin{ruledtabular}\begin{tabular}{lllll}
      Model & IF equal & IF unequal & H4 equal & H4 unequal \\
      \hline
      $q$                        & $1$     & $0.816$        & $1$     & $0.816$ \\
      $M_\mathrm{b}^\mathrm{tot}$ [M$_\odot$]  & $3.25$  & $3.25$         & $3.04$  & $3.04$ \\
      $M_\mathrm{b}$ [M$_\odot$]        & $1.63$  & $1.44, 1.81$   & $1.52$  & $1.35, 1.69$ \\
      $M_\mathrm{g}$ [M$_\odot$]        & $1.51$  & $1.36, 1.67$   & $1.40$  & $1.26,1.54$ \\
      $M_\mathrm{g}/R_c$                  & $0.140$ & $0.120, 0.164$ & $0.148$ & $0.132, 0.164$ \\
      $f_0$ [Hz]                 & $295$   & $234$          & $263$   & $263$ \\
      $d$ [km]                   & $59.3$  & $68.0$         & $61.0$  & $61.0$ \\
      $E_{B}$ [$10^{40}$erg]     & $8.19$  & $8.03$         & $9.51$  & $9.32$ \\
      $B_\mathrm{max}$ [$10^{12}$G]     & $1.99$  & $1.99$         & $1.99$  & $1.99$ \\
      $A_b$                     & $2.20$  & $0.76, 5.36$  & $1.97$  & $1.21, 3.13$
  \end{tabular}\end{ruledtabular}
   \label{tab:init_param}
\end{table}

\section{Initial Data \label{sec_initial_data}}

We evolve magnetized, quasicircular and irrotational BNS models. The
main properties of the initial data used for our simulations are
listed in Table~\ref{tab:init_param}. These data are produced using
the spectral-method code LORENE (http://www.lorene.obspm.fr), except
for the setup of the magnetic field (see below). We employ the
ideal-fluid EOS (denoted \texttt{IF} in the table) and the H4 EOS
(denoted \texttt{H4} \cite{Glendenning1991}), along with poloidal
initial magnetic fields that are confined inside the stars. The ideal-fluid 
EOS uses a polytropic index $\Gamma=2$ and a polytropic constant
$K=100$ as in previous
simulations~\cite{Giacomazzo2011PhRvD..83d4014G, Rezzolla:2011:6}. The
H4 EOS is instead implemented as a piecewise polytropic EOS as
described in Ref.~\cite{Read:2009:124032}. In order to also take 
thermal effects into account in this case, we add a thermal part via an ideal-fluid 
EOS with a polytropic index $\Gamma=1.8$ as done
in Refs.~\cite{Kiuchi:2014:41502, Kiuchi:2015:1509.09205}. The total masses
have been chosen so that the ideal-fluid and H4 equal-mass models are
the same as the ones evolved in Ref.~\cite{Rezzolla:2011:6} and
Refs.~\cite{Kiuchi:2014:41502, Kiuchi:2015:1509.09205}, respectively. All
our models inspiral for $\sim 3-6$ orbits before merger. Time of
merger is defined as the time of maximum amplitude in the GW signal.
 
For the ideal-fluid equal-mass simulations, we use three different magnetic field
orientations: both NS magnetic fields aligned to the orbital rotation
axis (\texttt{UU}), aligned and antialigned (\texttt{UD}), and both
antialigned (\texttt{DD}). For the ideal-fluid unequal-mass simulation, and
also for the H4 equal- and unequal-mass simulations, we use
the \texttt{UU} magnetic field configuration. In summary, there are six
models according to EOSs, mass ratio, and magnetic field
configurations: \texttt{IF\_q10\_UU},
\texttt{IF\_q10\_UD}, \texttt{IF\_q10\_DD}, \texttt{IF\_q08\_UU}, \texttt{H4\_q10\_UU}, and
\texttt{H4\_q08\_UU}. All the initial data computed with LORENE are publicly available online,
except for model \texttt{IF\_q10} (ideal-fluid equal-mass) which is
already available on the LORENE web page as
model \texttt{G2\_I14vs14\_D4R33\_45km}.

The magnetic fields are added a posteriori on top of the initial data
produced with LORENE using the following vector potential:

\begin{equation}
\label{eq:Avec}
A_{\phi} \equiv \varpi^2 A_b\, {\rm max}\,(p-p_{\rm cut},0)^{n_{\rm s}} \,,
\end{equation}

where $\varpi$ is the coordinate distance from the NS spin axis,
$p_{\rm cut}=0.04 \,\mathrm{max}(p)$ is a cutoff that determines where
the magnetic field goes to zero inside the NS, $\mathrm{max}(p)$ is
the initial maximum pressure in each star, and $n_{\rm s}=2$ is the
degree of differentiability of the magnetic field
strength~\cite{Giacomazzo2011PhRvD..83d4014G}. The values for $A_b$
for each model are listed in table~\ref{tab:init_param}. For the
unequal-mass models different values for $A_b$ were used for each star
in order to guarantee that they had the same initial magnetic field
strength. Antialigned fields are instead obtained by multiplying
$A_b$ by $-1$.

\begin{table}
  \caption{System properties for the different EOS and mass ratios considered in this work:
  BH mass ($M_\mathrm{BH}$), spin 
  ($a_\mathrm{BH}$), and disk mass ($M_\mathrm{disk}$) at the end of our simulations 
  ($27-30$ ms after collapse), accretion rate ($\dot{M}$), accretion timescale 
  ($\tau_\mathrm{acc}\equiv M_\mathrm{disk}/\dot{M}$), 
  time of BH formation since merger ($t_\mathrm{BH}$), instantaneous GW 
  frequency at merger ($f_\mathrm{merger}$) and characteristic GW 
  frequency in the HMNS phase ($f_\mathrm{HMNS}$). The accretion rate is taken as time 
  average from 5 ms after collapse to the end of the simulation. The time of merger 
  $t=0$ corresponds to the maximum GW strain. $f_\mathrm{HMNS}$
  is estimated from the characteristic peak in the post-merger spectrum 
  (see Section~\ref{sec_GWs}).} 
  \begin{ruledtabular}\begin{tabular}{lllll}
      Model & IF equal & IF unequal & H4 equal & H4 unequal \\
      \hline
      $M_\mathrm{BH}$ [M$_\odot$]  & $2.92$  & $2.78$         & $2.67$  & $2.50$ \\
      $a_\mathrm{BH}$        & $0.81$  & $0.77$   & $0.71$  & $0.63$ \\
      $M_\mathrm{disk}$ [M$_\odot$]     & $0.04$  & $0.21$   & $0.04$  & $0.23$ \\
      $\dot{M}$ [M$_\odot$/s]     & $0.8$ & $2.6$ & $1.1$ & $1.8$ \\
      $\tau_\mathrm{acc}$ [s]          & $0.05$   & $0.08$          & $0.03$   & $0.13$ \\
      $t_\mathrm{BH}$ [ms]        & $8.7$  & $1.3$   & $11.6$  & $24.7$ \\
      $f_\mathrm{merger}$ [kHz] & $1.36$ & $0.96$ & $1.43$ & $1.62$ \\
      $f_\mathrm{HMNS}$ [kHz] & $-$ & $-$ & $2.47$ & $2.69$
  \end{tabular}\end{ruledtabular}
   \label{tab:final-system}
\end{table}

\section{Evolution \label{evolution}}

\begin{figure*}
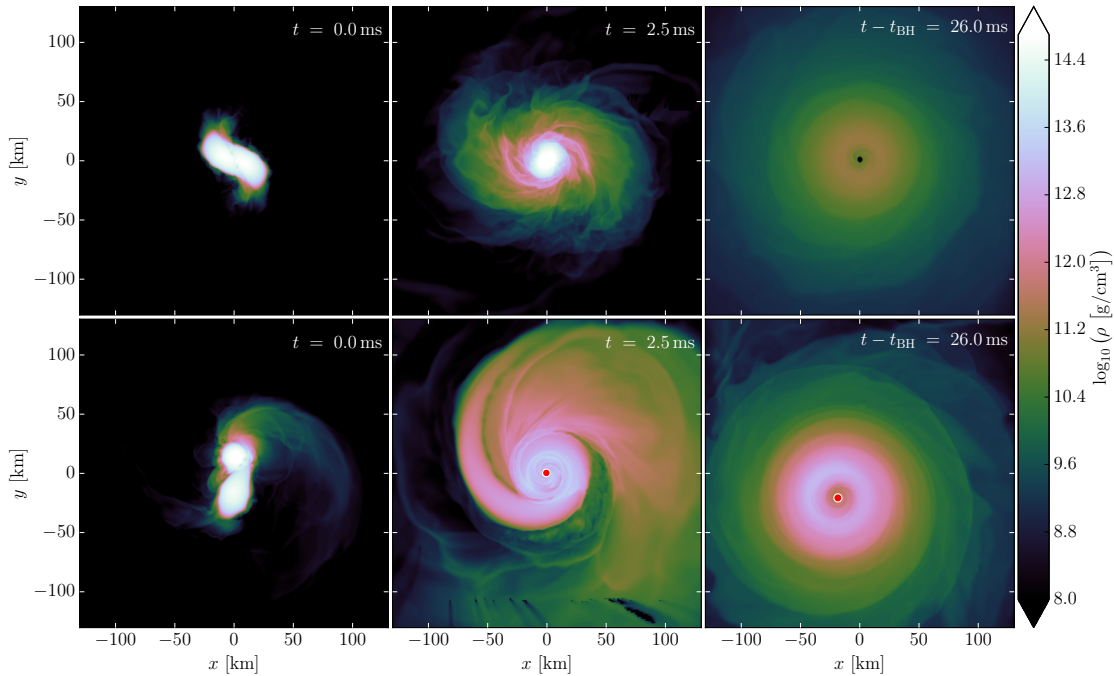

  \centering
  \includegraphics[width=0.99\textwidth]{{{cuts_snapsh_xy_IF}}}
  \caption{Rest-mass density evolution on the equatorial plane for models \texttt{IF\_q10} (top) 
  and \texttt{IF\_q08} (bottom). The horizon is marked with a red circle, with the exception of
  the top right panel which shows the excised region (black) instead.
  }
  \label{fig:2d_rms_cut_xy_IF}
\end{figure*}

\begin{figure*}
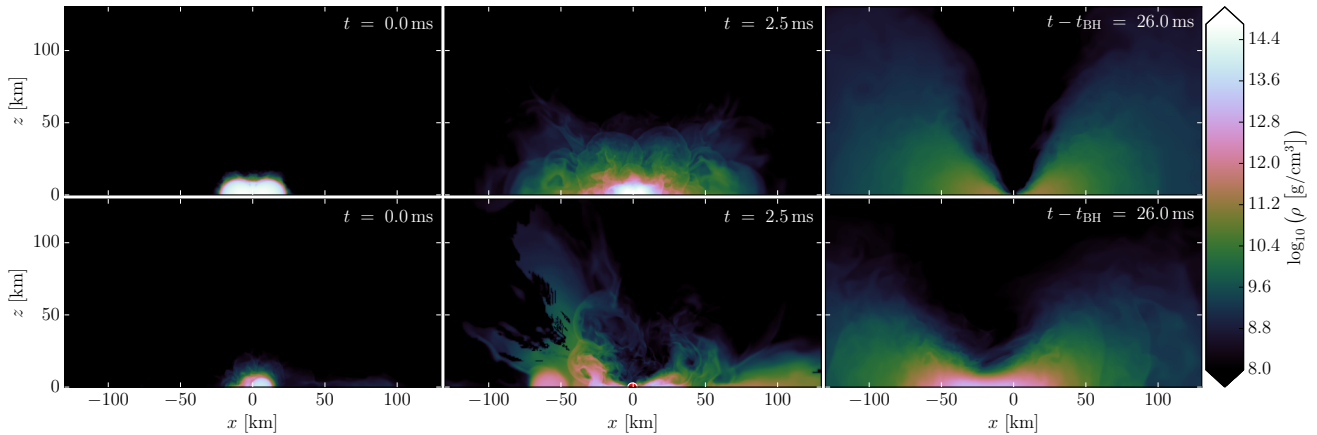

  \centering
  \includegraphics[width=0.99\textwidth]{{{cuts_snapsh_xz_IF}}}
  \caption{Rest-mass density evolution on the meridional plane for models \texttt{IF\_q10} (top) 
  and \texttt{IF\_q08} (bottom). Note the lower right panel constitutes an off-center cut because of 
  the BH drift.
  }
  \label{fig:2d_rms_cut_xz_IF}
\end{figure*}

In this section we provide an extensive discussion of the results of
our simulations, including the general dynamics, the magnetic field
evolution, the dependence on the EOS and the mass ratio, a
comparison with previous work, and a resolution study. The connection
to SGRBs and GW emission are discussed in
Secs.~\ref{sec_SGRB} and \ref{sec_GWs}.  Important
quantities characterizing the system are summarized in Table
\ref{tab:final-system} for the different cases considered in this
work.

\subsection{Ideal-Fluid Equal-Mass Model}
\label{sec:if_uu_q10}

We first consider the equal-mass case with ideal-fluid EOS and initial magnetic 
fields aligned with the orbital axis, \texttt{IF\_q10\_UU}. 
The following discussion refers to the standard resolution simulation, while different 
resolutions for this case are considered in Sec.~\ref{res}.

\begin{figure}
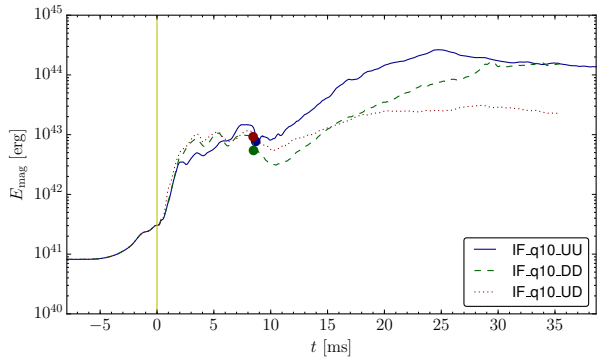

  \centering
  \includegraphics[width=0.95\columnwidth]{{{emag}}}
  \caption{Comparison of total magnetic energy between the models \texttt{IF\_q10\_UU}, \texttt{IF\_q10\_DD},
    \texttt{IF\_q10\_UD}. The yellow vertical line marks the merger time and the 
    circles show the time of BH formation for each model.}
  \label{fig:emag}
\end{figure}

\begin{figure}
  \centering
  \includegraphics[width=0.95\columnwidth]{{{Bmax}}}
  \caption{Comparison of the maximum values of magnetic field strength
    between the models \texttt{IF\_q10\_UU}, \texttt{IF\_q10\_DD}, \texttt{IF\_q10\_UD}.
    The yellow vertical line marks the merger time and the 
    circles show the time of BH formation for each model.}
  \label{fig:Bmax}
\end{figure}

\begin{figure}
  \centering
  \includegraphics[width=0.95\columnwidth]{{{Bmean}}}
  \caption{Comparison of the mean values of magnetic field strength
    between the models \texttt{IF\_q10\_UU}, \texttt{IF\_q10\_DD}, \texttt{IF\_q10\_UD}.
    The yellow vertical line marks the merger time and the 
    circles show the time of BH formation for each model.}
  \label{fig:Bmean}
\end{figure}

\begin{figure*}
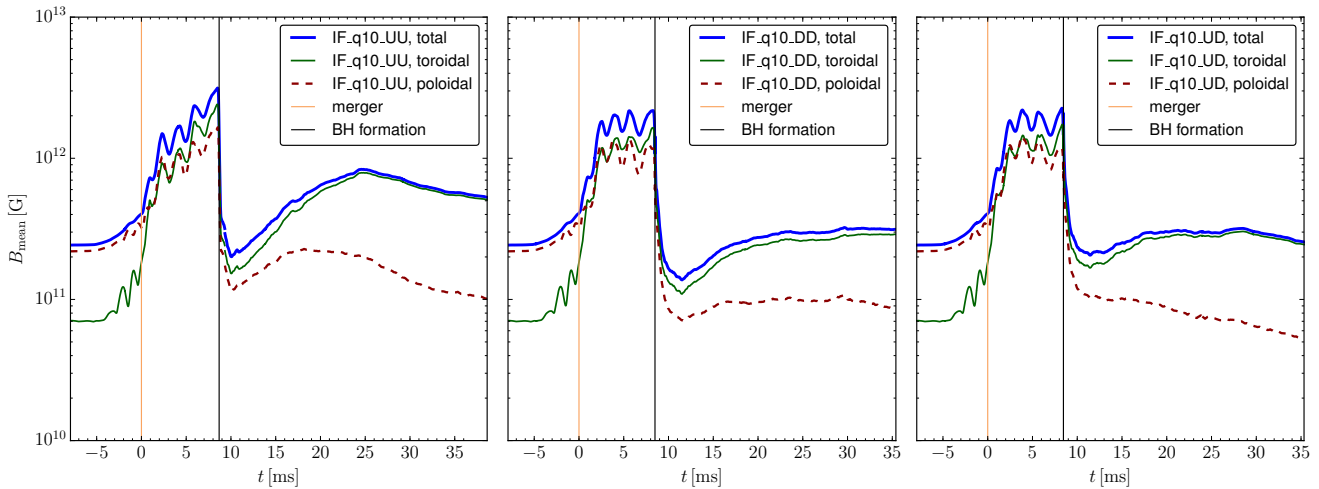

  \centering
  \includegraphics[width=0.99\textwidth]{{{Bmean_multisim}}}
  \caption{Comparison of the mean values of magnetic field strength
    between the models \texttt{IF\_q10\_UU}, \texttt{IF\_q10\_DD}, \texttt{IF\_q10\_UD}, including mean values 
    of toroidal and poloidal field components. The yellow vertical lines mark the merger time and 
    the black vertical lines mark the time of BH formation.}
  \label{fig:Bmean_multisim}
\end{figure*}

The rest-mass density evolutions on the equatorial and meridional planes
are shown in the top rows of Figures~\ref{fig:2d_rms_cut_xy_IF} and
\ref{fig:2d_rms_cut_xz_IF}, respectively. As its total
rest mass is well within the hypermassive regime for a single object,
the merger is followed by a HMNS phase lasting $\sim8.5$ ms and the
eventual collapse to a BH.  Most of the rest mass in the system is
rapidly swallowed by the BH during its formation, leaving behind only
a light disk. At the end of the simulation ($\sim$26 ms after BH
formation) the disk mass is only $\sim0.04$~M$_\odot$ and the
accretion time scale is less than 100 ms
(see Table~\ref{tab:final-system}).  The BH spin is relatively high
$a_\mathrm{BH}\sim 0.8$ (the highest value obtained in this study).

The evolutions of the magnetic field energy and strength are shown in Figures 
\ref{fig:emag}, \ref{fig:Bmax}, \ref{fig:Bmean}, and \ref{fig:Bmean_multisim}. 
A sudden increase of magnetic energy is observed in the first 2 ms after merger. This is 
to be attributed to the shear that is generated when the two stars first touch and that is associated 
with strong magnetic field amplification via the Kelvin-Helmholtz instability (although our 
resolution does not allow us to fully resolve it; see Section~\ref{res}).
In the following evolution, the magnetic field is further amplified (at a lower rate) 
in the HMNS phase and in the remnant disk after BH formation. The magnetic energy 
and the maximum field strength do not show a sudden decrease at BH formation, indicating 
that most of the field is outside the high-density bulk of the HMNS that is immediately 
swallowed by the nascent BH. Conversely, such a drop is observed when considering a 
density-weighted average of the magnetic field strength. 
Around 15 ms after BH formation the gain in magnetic energy becomes lower than the 
loss associated with the accretion of magnetized material in the disk. Overall, the maximum 
magnetic field strength achieved is a factor of $\sim50$ higher than the initial value.
More details on the magnetic field amplification mechanisms and the dependence on 
resolution are discussed in Section~\ref{res}.

As shown in Figure \ref{fig:Bmean_multisim}, magnetic field amplification is mostly in 
favor of the toroidal component. In terms of average magnetic field strength, the toroidal 
component becomes comparable to the poloidal one in the first ms after merger and in the 
HMNS phase the two keep growing together. Then, after BH formation the poloidal field remains 
much smaller than the toroidal one, which is more efficiently amplified in the disk.

We now discuss in more detail the geometrical structure of the magnetic field. 
To qualitatively assess the global structure of the field, we use three-dimensional
(3D) plots of selected field lines. Visualizing field lines is a complex task and can be very 
misleading. We developed a prescription for the automated selection of field lines that gives 
good results without any manual (i.e. potentially biased) intervention. The procedure is
described in detail in the Appendix. For a quantitative description of the field, we rely
instead on histograms of magnetic energy in suitable bins based on spatial position.

An overview of the evolution of the field structure is given in Fig.~\ref{fig:field3d_evol_IF_UU}.
During early inspiral, the field is given by the initial data prescription, Eq.~(\ref{eq:Avec}).
We recall that the magnetic field strength drops to zero towards the surface and there is no
field outside the stars.
During the last orbits of inspiral (not shown in the figure), the field already becomes more 
irregular. The complex fluid flows during merger finally destroy all regularity, as can be seen
in the second snapshot ($\sim2\usk\milli\second$ after merger).
In the remaining evolution, the field structure becomes more regular again. 
As expected, magnetic winding
produces a toroidal field of increasing strength near the equatorial plane.
More interestingly, we also observe a cone-like region of increasing strength along
the edge of the accretion torus. The alignment is highlighted in 
the figure by displaying two isodensity surfaces in addition to the field.
Initially, the field along the cone is more or less tangential, but still relatively
irregular. At a later stage, around $30\usk\milli\second$ after merger, the lines
along the cone acquire a clear ``twister'' structure.
This could be attributed to the stretching of field lines by the fluid flow along the edge of 
the torus.

By using an interactive version of Fig.~\ref{fig:field3d_evol_IF_UU} to look at magnified 
parts from different angles, we found that the strong field lines typically turn around sharply 
at some point and very closely follow their previous path in reverse. This is indeed
the expected outcome of stretching an initially irregular field  continuously along
a quasistationary shearing fluid flow.
We stress that Fig.~\ref{fig:field3d_evol_IF_UU} visualizes the orientation of the field, 
but not the sign, which alternates on small length scales. The cone contains field lines going 
both upwards and downwards (along the cone), and the toroidal field near the equatorial plane 
contains field lines wound both clockwise and counterclockwise.
The field near the BH axis is only mildly collimated. From animations showing a cut through 
the meridional plane, we found that it is also strongly fluctuating. This seems to be related 
to lumps of low-density matter falling towards the BH along the axis. 

To quantify the magnitude and topology of the magnetic field, we sum the magnetic field energy contained in bins regularly 
spaced in $\cos(\theta)$, where $\theta$ is the angle to the BH axis. Thus, a homogeneous 
field would result in a flat distribution. This measure allows us to distinguish 
the amount of energy in the disk, along the conical structure separating the disk and funnel, and near the axis. 
As a measure for the strength of the field, we computed for each bin the 
field strength $B_{90}$, defined by the requirement that 90\% of the magnetic field energy
is contributed by regions with field strength below $B_{90}$. We use this measure because
using the maximum field strength is too sensitive to potential outliers, while using the 
average field strength would depend on the volume under consideration. Using $B_{90}$ is a 
good compromise.

\begin{figure*}
  \centering
  \includegraphics[width=0.99\textwidth]{{{bfield_dens_3d_IF_UU_evol}}}
  \caption{Evolution of the magnetic field structure for model \texttt{IF\_q10\_UU}. 
  Top left: inspiral phase, showing the magnetic field, as well as the lower half of the NS surfaces.
  Top center: magnetic field $2\usk\milli\second$ after merger together with the isodensity surface for 
  $5\times 10^{12}\usk\gram\per\centi\meter\cubed$, drawn as a semitransparent red surface. 
  Top right: magnetic field structure $12\usk\milli\second$ after merger.
  Bottom left: magnetic field $22\usk\milli\second$ after merger, together with two isosurfaces
  of density $10^8$ (yellow) and $10^{10}\usk\gram\per\centi\meter\cubed$ (cyan), cut off for $y<0$.
  Bottom right: same at $35\usk\milli\second$ after merger. The color of the field lines gives a 
  rough indication of the field strength (see colorbar), but for quantitative results compare 
  figures~\ref{fig:hist_evol_IF_UU}, \ref{fig:hist_final_IF_all}, and \ref{fig:hist_final_H4_IF}. 
  The procedure for selecting which field lines to plot is described in the appendix.}
  \label{fig:field3d_evol_IF_UU}
\end{figure*}

\begin{figure}
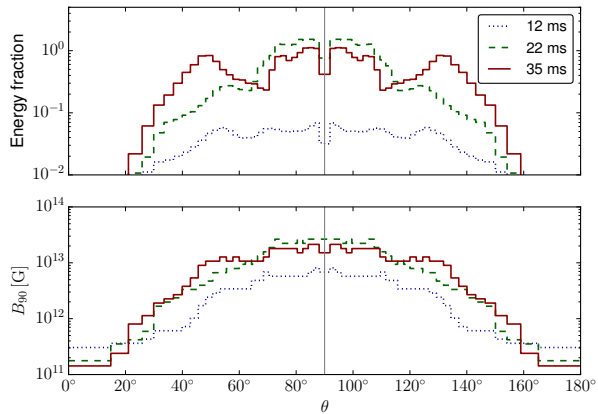

  \centering
  \includegraphics[width=0.95\columnwidth]{{{hist_evol_IF_UU}}}
  \caption{Distribution of the magnetic field with respect to the $\theta$-coordinate, 
  for model \texttt{IF\_q10\_UU} at various times after merger.
  Top: histogram of magnetic energy employing bins regularly spaced in
  $\cos(\theta)$, where $\theta=0$ is the z-axis and $\theta=90\degree$ the equator. 
  The plot is normalized to the total magnetic energy 
  $35\usk\milli\second$ after merger.
  Bottom: field strength $B_{90}$ defined as the value for which 90\% of the 
  magnetic energy (inside a given $\cos(\theta)$ bin) is contained in regions 
  with field strength below $B_{90}$.}
  \label{fig:hist_evol_IF_UU}
\end{figure}

The energy distribution and the field strength $B_{90}$ for model 
\texttt{IF\_q10\_UU} at three different times are shown in Fig.~\ref{fig:hist_evol_IF_UU}.
The total magnetic energy near the equatorial plane increases by around an order of 
magnitude between $12$--$22\usk\milli\second$ after merger, most likely
because of magnetic winding in the torus. The energy $35\usk\milli\second$ after merger is slightly lower,
however. The reason is uncertain, but it might be a change of the torus structure and/or
loss by accretion. The energy along the conical structure separating the disk and funnel is steadily 
growing (side peaks). The final distribution has a 
pronounced local maximum, corresponding to an opening half-angle around $50\degree$. 
Notably, the regions near the BH axis ($\theta<20\degree$) do not contribute significantly 
to the total field energy.

The field strength $B_{90}$ near the equator increases from ${\approx}6\times 10^{12}\usk\Gauss$
at $12\usk\milli\second$ after merger up to ${\approx}2\times 10^{13}\usk\Gauss$ at
$22\usk\milli\second$ after merger, and afterwards it stagnates.
$B_{90}$ is of the same order of magnitude at all angles from the equator up to the 
conical structure, and then it drops rapidly in the funnel.
In particular, near the axis the field is very weak (less than $3\times 10^{11}\usk\Gauss$ 
at $12\usk\milli\second$ after merger) and further drops by a factor of ${\approx}2$ at the end of the 
simulation.

\subsection{Comparison with Rezzolla et al. 2011}
\label{sec:cp-mising-link}

As mentioned before, the specific choice of EOSs used in this work has
been made in order to favor comparisons with previous work. In
particular, our equal-mass model employing the simplistic ideal-fluid EOS 
is the same as the one studied in 
\cite{Rezzolla:2011:6}, the first work to claim the formation 
of a funnel-like structure in the magnetic field after
BH formation, a region of low-density
matter where a jet eventually producing a GRB may be launched. 

In order to make a meaningful comparison between the present work and
Ref.~\cite{Rezzolla:2011:6}, we first describe the differences in
the numerical methodology of the simulations. However, we did not investigate 
the influence of different parameters one by one because it would have been 
too expensive. Below we report what we believe are the relevant changes.

First of all, in both works the vector potential is the evolved variable
for the magnetic field, in order to guarantee the divergence-free character 
of the magnetic field. However, differently from Ref.~\cite{Rezzolla:2011:6}, we use the 
modified Lorenz gauge \cite{Etienne:2011re, Farris2012}. 
This avoids spurious amplifications of the magnetic field at the boundary 
between refinement levels, as was observed in the simulations of 
Ref.~\cite{Rezzolla:2011:6}.

The resolution of the simulation in Ref.~\cite{Rezzolla:2011:6} is the same
as our standard resolution, as is the number of refinement levels. 
In the current work, we evolved the same model also with higher and lower
resolutions, as discussed in Sec.~\ref{res}.
The location of the outer boundary and the size of the refinement levels
are different from Ref.~\cite{Rezzolla:2011:6}.
The finest refinement level after merger in this work only extends to 
$30 \usk\kilo\meter$, compared to $44\usk\kilo\meter$ used in
Ref.~\cite{Rezzolla:2011:6}. The outer boundary on the other hand was expanded 
to $1403 \usk\kilo\meter$, almost 4 times the extent used in 
Ref.~\cite{Rezzolla:2011:6}.
We believe that this was an important improvement on the previous work. 
The simulation described in Ref.~\cite{Rezzolla:2011:6} had to be terminated
when large spurious waves in the magnetic field coming from the outer
boundary had contaminated the solution even near the central object, while
we encountered no such problems.

Another difference concerns the symmetries. In both works, a
reflection symmetry with respect to the orbital plane was used, but in
contrast to Ref.~\cite{Rezzolla:2011:6}, we do not enforce $\pi$ symmetry
around the $z$ axis, thus allowing for non-$\pi$-symmetric modes to
develop. However, in the case of equal-mass binaries the system
becomes roughly axisymmetric soon after the merger and therefore we do
not think that the different symmetries imposed led to significant
differences in the results.

Another improvement is the lower density of the artificial atmosphere
in our work, $\sim 6.2\times 10^4 \usk\gram\per\centi\meter\cubed$,
which is 3 orders of magnitude smaller than the one used in
Ref.~\cite{Rezzolla:2011:6}.  This could be relevant for the computation of
the accretion rate, estimated in Ref.~\cite{Rezzolla:2011:6} from the time
derivative of the total amount of matter outside the apparent horizon,
and which might contain a significant error due to the effect of the
artificial atmosphere. We measure the accretion rate from the
integrated matter flux through the apparent horizon instead.

We now compare the outcome of Ref.~\cite{Rezzolla:2011:6} to our standard resolution 
run of the same model. The most important 
improvement is our detailed analysis of the magnetic field near the BH spin axis.
In Ref.~\cite{Rezzolla:2011:6}, a magnetic field of $8\times 10^{14} \usk\Gauss$
near the axis\footnote{However, L.B. and B.G. (who are also co-authors 
of Ref.~\cite{Rezzolla:2011:6}) found this to be an erroneous statement.
The number quoted in \cite{Rezzolla:2011:6} referred to the \emph{global} 
maximum of the poloidal field component (see also figure 2 of~\cite{Rezzolla:2011:6}).} 
was reported. In this work, we found a much weaker field near 
the axis. In fact, we computed the full magnetic field energy spectrum as a function
of the angle to the spin axis, and found that 90\% of the field energy near 
the axis (cf. Fig.~\ref{fig:hist_evol_IF_UU}) is contributed by field 
strengths below $2\times 10^{11} \usk\Gauss$, and that the spectrum does not 
extend beyond $10^{12} \usk\Gauss$.

Further, we find only a weakly collimated 
and fluctuating field in this region. We could not reproduce the strong 
collimation suggested by the field line visualization of Fig.~3 in 
Ref.~\cite{Rezzolla:2011:6}, which shows field lines originating on the apparent 
horizon and tracing the shape of the funnel, proceeding outwards nearly as 
straight lines. One could argue that this is merely a difference in 
visualization methods, given that the seeds of this plot were selected 
ad hoc, while we adopted a more systematic approach (see the Appendix)
for the selection of field lines. However, we do not fully rely on such visualizations 
and also used two-dimensional (2D) cuts in the meridional plane, both as snapshots and 
animations, to cross-check our results.
What we find instead is a twister-like configuration of the magnetic field,
with an opening half-angle around $50\degree$ and a field strength around 
$10^{13} \usk\Gauss$.

Comparing the evolution of the maximum field strength, i.e.,
Fig.~\ref{fig:Bmax} with the right panel of Fig.~2 in Ref.~\cite{Rezzolla:2011:6},
we find a slightly stronger amplification between merger and collapse. The main 
difference however is the post-collapse amplification.
The maximum field strength in Ref.~\cite{Rezzolla:2011:6} keeps growing up to 
$10^{15}\usk\Gauss$, while for our simulation it settles around $10^{14}\usk\Gauss$.
Also, our simulation is a bit longer and exhibits a decrease of the maximum field 
strength starting $24\usk\milli\second$ after merger.
These differences may be due to the different numerical
setups of the two simulations, in particular the location of the outer
boundary, but we cannot provide certain conclusions. 

\begin{figure*}
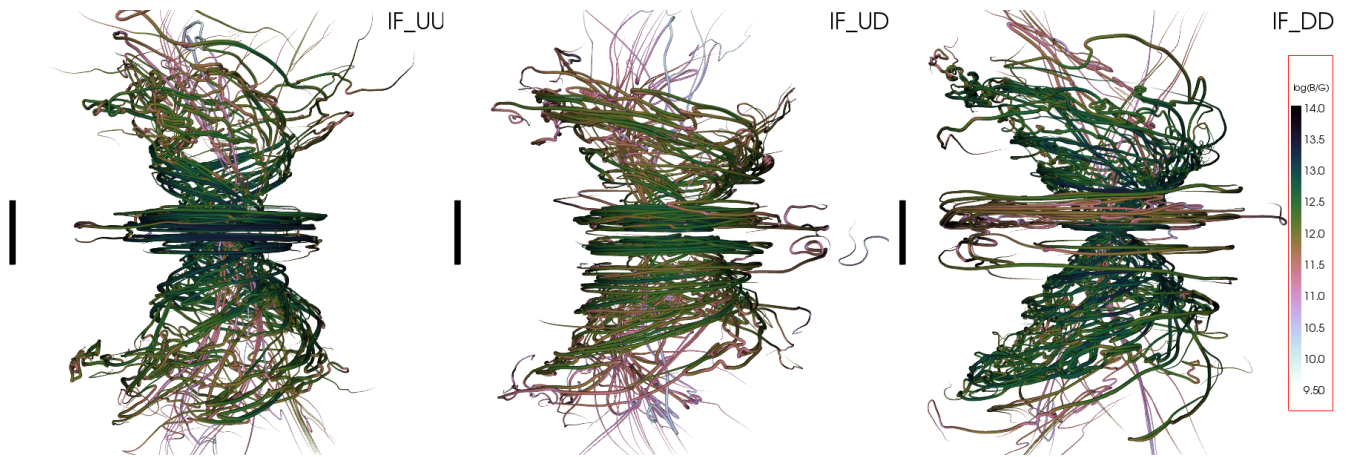

  \centering
  \includegraphics[width=0.99\textwidth]{{{field3d_final_IF_all}}}
  \caption{Magnetic field structure $35 \usk\milli\second$ after the merger,
  comparing models \texttt{IF\_q10\_UU}, \texttt{IF\_q10\_UD}, and \texttt{IF\_q10\_DD}. 
  The black bars provide a length scale of $20 \usk\kilo\meter$.
  The coloring of the fieldlines indicates the magnetic 
  field strength ($\log_{10}(B\,[\mathrm{G}])$, same colorscale for all models) along 
  the lines. However, for quantitative results see Fig.~\ref{fig:hist_final_IF_all}. }
  \label{fig:field3d_final_IF_all}
\end{figure*}

\begin{figure}
  \centering
  \includegraphics[width=0.95\columnwidth]{{{hist_final_IF_all}}}
  \caption{Like Fig.~\ref{fig:hist_evol_IF_UU}, but comparing models 
  \texttt{IF\_q10\_UU}, \texttt{IF\_q10\_UD}, and \texttt{IF\_q10\_DD}
  $35 \usk\milli\second$ after the merger. The energy distribution (top 
  panel) is normalized to the total energy for model \texttt{IF\_q10\_UU}.
  }
  \label{fig:hist_final_IF_all}
\end{figure}

We stress that the maximum 
is not a very reliable measure for the growth of the magnetic field,
since it is sensitive to outliers, either physical or caused by numerical errors. 
Inspecting measures not relying on a single point is more meaningful.
In particular, the measure $B_{90}$ is a more robust replacement for the maximum.
Furthermore, using the density-weighted mean allowed us to quantify the field of the 
HMNS (see Fig.~\ref{fig:Bmean}). More specifically, the use of histograms of magnetic 
energy with respect to the $\theta$ coordinate allowed us to quantify
the spatial distribution of the post-collapse field in more detail (see 
Fig.~\ref{fig:hist_evol_IF_UU}). As in Ref.~\cite{Rezzolla:2011:6}, we find
a clearly toroidal field structure in the disk, although the maximum 
strength is more than 1 order of magnitude lower than the value 
$2\times 10^{15} \usk\Gauss$ reported in Ref.~\cite{Rezzolla:2011:6}.
Further, the measure $B_{90}$ is around 2 orders of magnitude lower.

Note that a comparison between our Fig.~\ref{fig:emag} and the
left panel of Fig.~2 of \cite{Rezzolla:2011:6} is not possible
because they show different quantities: the former shows the total
magnetic energy as integrated over the whole domain, while the latter
shows the emitted magnetic energy computed by integrating the Poynting
vector. We did not compute the latter in our simulation.

The mass and spin we found for the BH formed during merger agree better than 
1\% with Ref.~\cite{Rezzolla:2011:6}. Also the initial disk mass is comparable.
We did however find an accretion rate around 4 times larger than the one 
reported in Ref.~\cite{Rezzolla:2011:6}. We believe our result is more robust since we
use the flux instead of the total rest mass outside the 
horizon, which in fact starts \emph{increasing} at some point for the data 
on which Ref.~\cite{Rezzolla:2011:6} is based.

Both Ref.~\cite{Rezzolla:2011:6} and the present work do not find any
outflows in the funnel along the rotation axis of the BH.
This might be due to missing physical input (neutrino treatment;
limits of the MHD approximation) in the simulations and/or too low
resolution.  We have checked that the matter in the funnel is not
magnetically dominated in our simulation, which makes outflows
unlikely.  We note that the simulations presented
in Refs.~\cite{Ruiz2016,Paschalidis:2015:14} featured mildly relativistic
outflows. This is due to the use of stronger initial magnetic fields
that allow to better resolve the magnetorotational instability (MRI), and much longer evolutions after
BH formation.  Finally, Ref.~\cite{Rezzolla:2011:6} reported some outflows
along the edge of the funnel.  However, the given limit
$\Gamma \lesssim 4$ for the Lorentz factor of the outflows was based
on the \emph{global} maximum.  Using a movie showing a cut of $v^z$ in
the x-z plane ($z>0$), we find a much lower limit of $v^z < 0.3 \usk c$
for any upward movement of matter in the disk or its edge.

\subsection{Effects of the Initial Magnetic Field Orientation}

When considering a different orientation for the initial magnetic field in the 
two NSs, we observe almost no differences in the overall dynamics, as well as the final BH mass 
and spin, the time of BH formation, the mass in the disk and the accretion rate.
Nevertheless, some differences can be observed in the magnetic field evolution. 
From the magnetic energy and the maximum field strength (Fig.  \ref{fig:emag} 
and \ref{fig:Bmax}) we see that the totally aligned (with respect to the orbital axis) 
or totally misaligned cases, \texttt{UU} and \texttt{DD} respectively, reach the same level of 
magnetic field amplification at the end of the simulation (although with a slightly 
different path). The case in which magnetic fields are aligned in one NS and 
antialigned in the other (\texttt{UD}) is instead disfavoured because of a less 
efficient amplification in the disk, after BH formation.
From the density-weighted average of the magnetic field strength 
(cf. Fig.~\ref{fig:Bmean}), we notice a stronger magnetic field amplification in 
the inner (highest-density) region of the accretion disk for the \texttt{UU} case, 
compared to the \texttt{DD} and \texttt{UD} cases.

The influence of the initial alignment on the final structure of the field 
is shown in Fig.~\ref{fig:field3d_final_IF_all}. All models exhibit the same
general features, namely a toroidal field near the equatorial plane, a 
twister-shaped field forming a conical structure, and a very weak field 
near the axis. The relative strength between the cone and equatorial parts
seems strongly affected by the initial alignment. This impression is validated
by Fig.~\ref{fig:hist_final_IF_all}, which shows the distribution of the magnetic 
energy and the field strength $B_{90}$ introduced in Sec.~\ref{sec:if_uu_q10}.
The \texttt{UU} configuration contains more energy near the equatorial plane than
both the \texttt{UD} and \texttt{DD} configurations, which are comparable in that respect.
The amount of energy in the cone, on the other hand, is largest for the \texttt{DD} 
case and smallest for the \texttt{UD} case. The latter also has the weakest field
strength $B_{90}$.

\subsection{Ideal-Fluid Unequal-Mass Model}

\begin{figure}
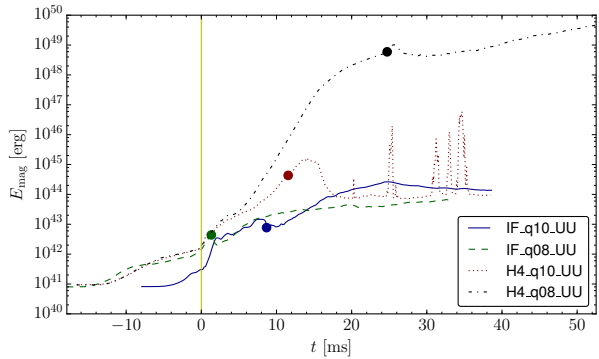

  \centering
  \includegraphics[width=0.95\columnwidth]{{{emag_eos_q}}}
  \caption{Comparison of total magnetic energy between models \texttt{IF\_q10\_UU}, \texttt{IF\_q08\_UU},
    \texttt{H4\_q10\_UU}, \texttt{H4\_q08\_UU}. The yellow vertical line marks the merger time and the 
    circles show the time of BH formation for each model.}
  \label{fig:emag_eos_q}
\end{figure}

\begin{figure}
  \centering
  \includegraphics[width=0.95\columnwidth]{{{Bmax_eos_q}}}
  \caption{Comparison of the maximum values of magnetic field strength
    between models \texttt{IF\_q10\_UU}, \texttt{IF\_q08\_UU}, \texttt{H4\_q10\_UU}, \texttt{H4\_q08\_UU}. 
    The yellow vertical line marks the merger time and the 
    circles show the time of BH formation for each model.}
  \label{fig:Bmax_eos_q}
\end{figure}

\begin{figure}
  \centering
  \includegraphics[width=0.95\columnwidth]{{{Bmean_eos_q}}}
  \caption{Comparison of the mean values of magnetic field strength
    between models \texttt{IF\_q10\_UU}, \texttt{IF\_q08\_UU}, \texttt{H4\_q10\_UU}, \texttt{H4\_q08\_UU}. 
    The yellow vertical line marks the merger time and the 
    circles show the time of BH formation for each model.}
  \label{fig:Bmean_eos_q}
\end{figure}

\begin{figure*}
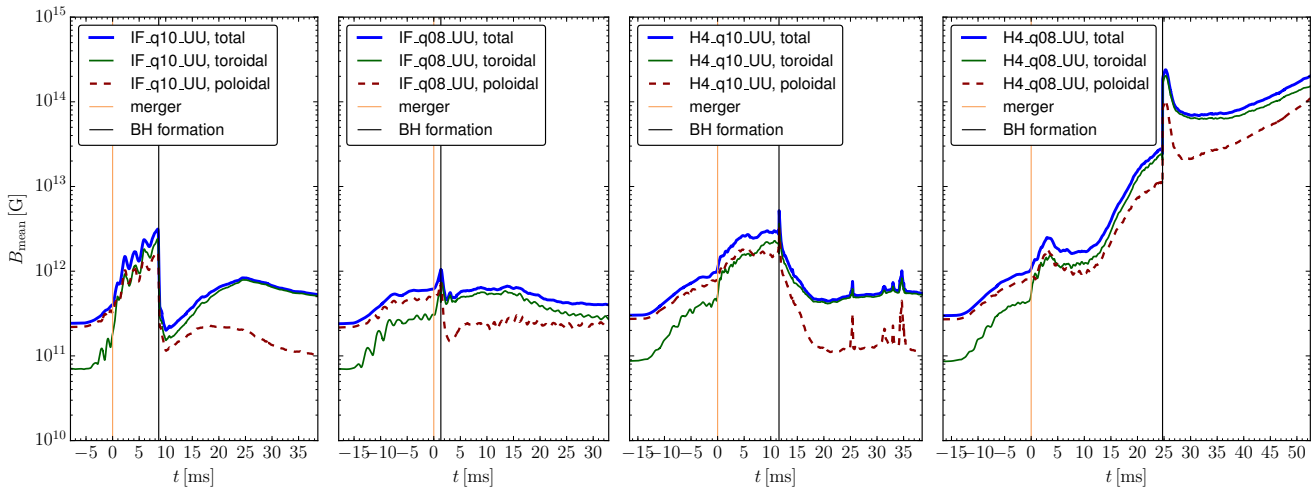

  \centering
  \includegraphics[width=0.99\textwidth]{{{Bmean_3mag_eos_q}}}
  \caption{Comparison of the mean values of the total, poloidal, and toroidal magnetic field strengths
    between models \texttt{IF\_q10\_UU}, \texttt{IF\_q08\_UU}, \texttt{H4\_q10\_UU}, \texttt{H4\_q08\_UU}.
  The yellow vertical lines mark the merger time and the black vertical
  lines mark the time of BH formation.}
  \label{fig:Bmean_3mag_eos_q}
\end{figure*}

In order to investigate the effect of the mass ratio on the dynamics of
matter and magnetic fields, we also evolved a model with a mass ratio
of $\sim 0.8$ (model \texttt{IF\_q08}).

The bottom rows of Figures~\ref{fig:2d_rms_cut_xy_IF}
and~\ref{fig:2d_rms_cut_xz_IF} show the evolution of the rest-mass
density on the equatorial and meridional planes, respectively.\footnote{In 
the central lower panel of both Figures~\ref{fig:2d_rms_cut_xy_IF} and
\ref{fig:2d_rms_cut_xz_IF} one can notice some artificial effects on the 
boundary between refinement levels, caused by failures in the 
conservative-to-primitive routine that sets those grid points to atmosphere. 
These effects, however, are present only in this case and they have negligible 
effect on the results discussed in this work. }
In this case the evolution is strongly asymmetric with the less compact star
being strongly deformed and disrupted during merger. Even if this
model has the same total baryonic mass as the equal-mass case, it
promptly forms a BH after merger and therefore does not produce a
HMNS. It is already evident from Figure~\ref{fig:2d_rms_cut_xy_IF}
that the disk formed after merger has higher densities and it is more
extended. As expected it is indeed more massive than the one formed in
the equal-mass case and it has a rest mass of $\sim0.21 M_\odot$ at
the end of the simulation. The accretion rate is more than $3$
times larger than in the equal-mass case, while the BH has a smaller
mass and spin (see Table~\ref{tab:final-system}), due to the larger
amount of mass still in the disk by the end of the simulation.

The evolution of the magnetic field strength is shown in
Figures~\ref{fig:emag_eos_q}, \ref{fig:Bmax_eos_q},
\ref{fig:Bmean_eos_q}, and~\ref{fig:Bmean_3mag_eos_q}. Because of the
lack of a HMNS phase, the magnetic field is not amplified to the same
maximum strengths as the equal-mass model prior to collapse, but, also
because of the fact that more mass is left outside the BH, the
density-weighted mean value after BH formation is similar to the
equal-mass model (compare the first and second panels of
Figure~\ref{fig:Bmean_3mag_eos_q}).

\begin{figure}
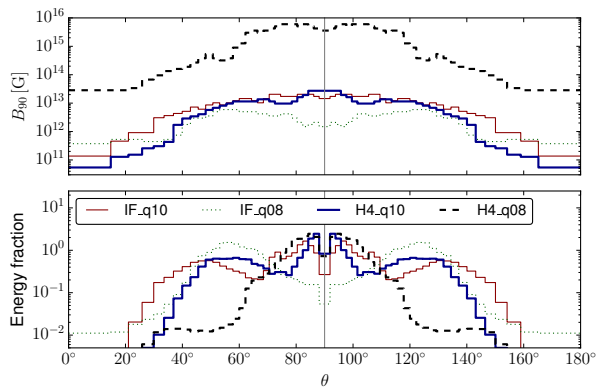

  \centering
  \includegraphics[width=0.95\columnwidth]{{{hist_final_H4_IF_b}}}
  \caption{Like Fig.~\ref{fig:hist_evol_IF_UU}, but comparing models 
  \texttt{IF\_q10\_UU}, \texttt{IF\_q08}, \texttt{H4\_q10}, and \texttt{H4\_q08} 
  around $32 \usk\milli\second$ after the merger. Also, each model is normalized
  separately in the lower panel, and
  we employed coordinates where the BH is located at the origin to account for the 
  BH drift exhibited by the unequal mass models.}
  \label{fig:hist_final_H4_IF}
\end{figure}

The influence of the mass ratio on the structure of the magnetic field
is shown in Figures~\ref{fig:hist_final_H4_IF}
and~\ref{fig:field3d_final_H4_IF}.  For the ideal-fluid models, we
find that the magnetic energy near the equatorial plane is reduced by
an order of magnitude for the unequal-mass case.  The energy and field
strength $B_{90}$ in the conical structure are comparable, but the opening
half-angle is ${\approx}10\degree$ larger for the unequal-mass case.
Note that we find much larger differences for the H4 EOS, as will be
discussed in Sec.~\ref{sec:h4_q08}.

\begin{figure*}
  \centering
  \includegraphics[width=0.99\textwidth]{{{field3d_final_H4_IF}}}
  \caption{Magnetic field structure around $32 \usk\milli\second$ after the merger,
  comparing models \texttt{IF\_q10\_UU}, \texttt{IF\_q08}, \texttt{H4\_q10}, 
  and \texttt{H4\_q08}. The black bars provide a length scale of $20 \usk\kilo\meter$.
  The coloring of the field lines indicates the magnetic 
  field strength ($\log_{10}(B\,[\mathrm{G}])$, same color scale for all models) along 
  the lines. However, for quantitative results see Fig.~\ref{fig:hist_final_H4_IF}.}
  \label{fig:field3d_final_H4_IF}
\end{figure*}

\subsection{Equal-Mass H4 Model}

We now investigate the effect of a different EOS using the piecewise
approximation of the H4 EOS. We begin by describing our equal-mass model,
which we recall is also the same one evolved in 
Refs.~\cite{Kiuchi:2014:41502,Kiuchi:2015:1509.09205}.

\begin{figure*}
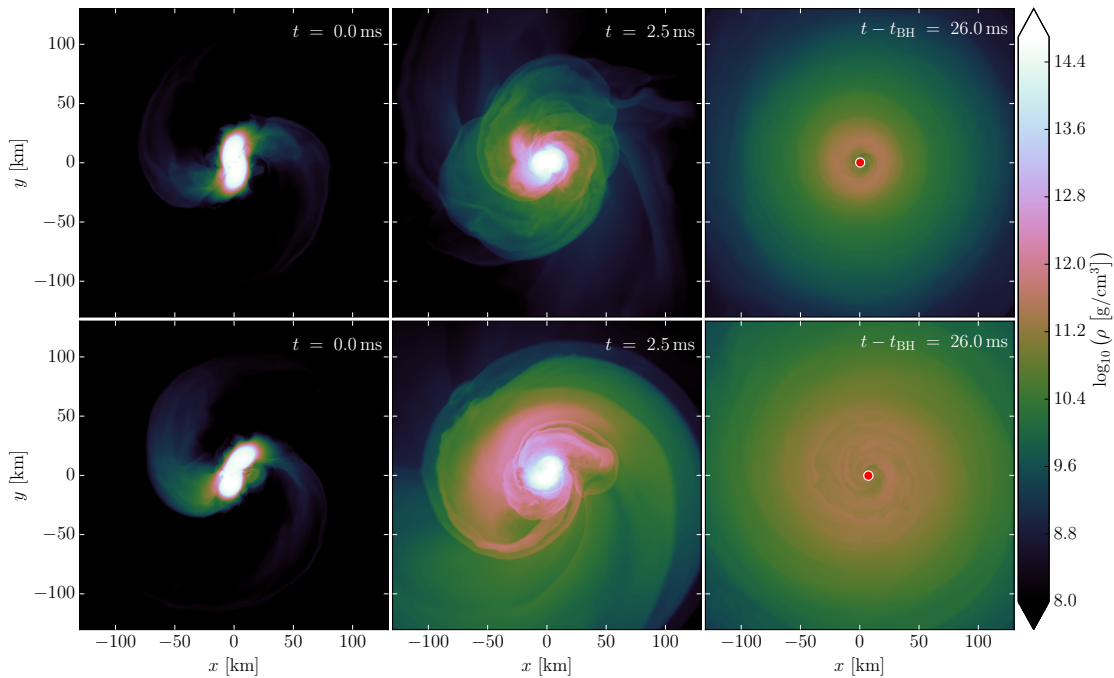

  \centering
  \includegraphics[width=0.99\textwidth]{{{cuts_snapsh_xy_H4}}}
  \caption{Rest-mass density evolution on the equatorial plane for models \texttt{H4\_q10} (top) and \texttt{H4\_q08} (bottom).}
  \label{fig:2d_rms_cut_xy_H4}
\end{figure*}

\begin{figure*}
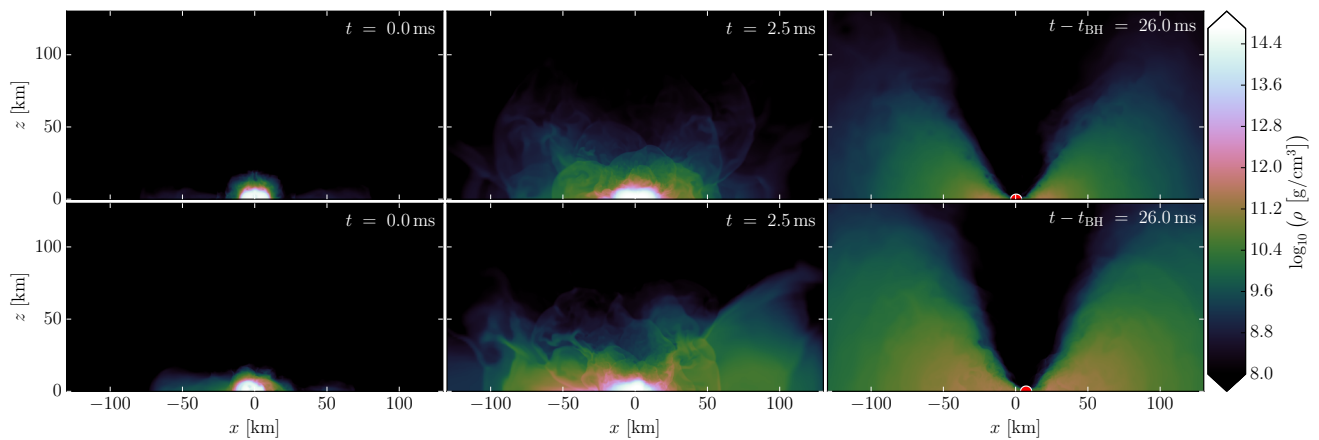

  \centering
  \includegraphics[width=0.99\textwidth]{{{cuts_snapsh_xz_H4}}}
  \caption{Rest-mass density evolution on the meridional plane for models \texttt{H4\_q10} (top) and \texttt{H4\_q08} (bottom).}
  \label{fig:2d_rms_cut_xz_H4}
\end{figure*}

The top panels of figures~\ref{fig:2d_rms_cut_xy_H4}
and~\ref{fig:2d_rms_cut_xz_H4} show the evolution of the rest-mass
density on the equatorial and meridional planes, respectively. Like in
the case of the ideal-fluid equal-mass model \texttt{IF\_q10\_UU}, the
merger remnant goes through a HMNS phase lasting about $12$ ms
before collapsing to a spinning BH. The disk mass is approximately
the same as in model \texttt{IF\_q10\_UU}, but the BH mass is slightly
smaller, consistent with the lower initial mass for the H4 models (see
Table~\ref{tab:final-system}).

The comparison of the magnetic field evolution between the H4 and the ideal-fluid
equal-mass models is shown in
Figures~\ref{fig:emag_eos_q}, \ref{fig:Bmax_eos_q}, \ref{fig:Bmean_eos_q},
and~\ref{fig:Bmean_3mag_eos_q}. Since the lifetime of the HMNS is slightly
longer than that of the ideal-fluid equal-mass model, the amplification of the
magnetic energy and the maximum field strength are larger than
in the ideal-fluid equal-mass model during the HMNS phase. After BH formation
the magnetic field in the disk has a strength comparable to the one
for the ideal-fluid equal-mass model, even if it exhibits a smaller decrease at
BH formation. This may also be correlated with the slightly higher
densities in the disk (compare the rightmost top panels of
Figures~\ref{fig:2d_rms_cut_xy_H4} and~\ref{fig:2d_rms_cut_xy_IF}).

In Figure~\ref{fig:emag_eos_q} one can also notice some spikes in the
evolution of the magnetic energy. These are due to very brief
amplifications of the magnetic field near the surface of the apparent
horizon in matter infalling into the BH and are very rapidly accreted
by the BH.

A comparison of the magnetic field structure for
models \texttt{H4\_q10} and \texttt{IF\_q10} is given in
Figures~\ref{fig:hist_final_H4_IF} and~\ref{fig:field3d_final_H4_IF}.
Note however that the masses of the stars are also different, not just
the EOS. The main difference is the opening half-angle of the conical
part of the field, which is ${\approx}10\degree$ larger for the H4
equal-mass case. The magnetic energy and field strength $B_{90}$ are
instead very similar (see Figure~\ref{fig:hist_final_H4_IF}).

\subsection{Comparison with Kiuchi et al 2014}
\label{sec:kiuchi}

Our equal-mass H4 EOS model allows for a direct comparison with the
results of Refs.~\cite{Kiuchi:2014:41502,Kiuchi:2015:1509.09205}, who studied
magnetized binaries with the highest grid resolution to date. 
For this, they employed a fixed mesh-refinement code described in 
Refs.~\cite{Kiuchi:2011:532,Kiuchi:2012:86}. The implementation of their 
fixed mesh refinement (except for the part dealing with the magnetic field) 
is based on that of the SACRA code \cite{Yamamoto:2008:78}, which had been 
quantitatively compared to the {\tt Whisky} code 
\cite{Baiotti:2003:210, Baiotti:2005:24035} 
several years ago in Refs.~\cite{Baiotti:2010:82, Read:2013:88}.
The main difference between {\tt Whisky} and the latest code of 
Refs.~\cite{Kiuchi:2014:41502, Kiuchi:2015:1509.09205} is the scheme used
to enforce the divergence-free constraint for the magnetic field. 
Differently from {\tt Whisky}, the code of 
Refs.~\cite{Kiuchi:2014:41502, Kiuchi:2015:1509.09205} employs a 
fourth-order-accurate-in-time flux-CT scheme \cite{Balsara:1999:149}, 
which ensures also the magnetic-flux conservation across refinement boundaries,
in addition to the divergence-free condition. Another difference is that
the artificial atmosphere density is only constant up to some fixed radius
and then falls of like $r^{-2}$ \cite{Kiuchi:2012:86}. This is important for 
ejected matter and magnetically driven winds, but probably irrelevant for 
the results discussed here.

The most important difference to the simulations presented in
Ref.~\cite{Kiuchi:2014:41502} is the grid resolution.  The finest grid
spacing used in Ref.~\cite{Kiuchi:2014:41502} is $70 \usk\meter$, which is
$2.66$ times better than our standard resolution. The extent of the
finest level is also larger than ours.  The outer boundary in our work
is slightly farther out than that in Ref.~\cite{Kiuchi:2014:41502}, but this is
probably scarcely relevant for the results discussed here.  In both
cases the computational domain should be large enough to allow the
evolution of the remnant and disk without the influence of boundary
effects.

For the equal-mass H4 model, we also performed a simulation with the same 
grid spacing of $150 \usk\meter$ used for the lowest-resolution 
runs in Ref.~\cite{Kiuchi:2014:41502}. In the following,
we compare our main results to the $150 \usk\meter$ resolution run in Ref.~\cite{Kiuchi:2014:41502} 
with the smallest initial magnetic field, $10^{15}\usk\Gauss$, which is 
still 500 times stronger than ours.
The strong field in Ref.~\cite{Kiuchi:2014:41502} was chosen to facilitate the 
study of magnetic instabilities, while our aim is to use values more likely
to occur in nature.

We find a HMNS lifetime of $10.9 \usk \milli\second$, which agrees within 
10\% with the value shown in Fig.~2 of Ref.~\cite{Kiuchi:2014:41502}. The dimensionless 
BH spin $10\usk\milli\second$ after merger in our simulation is $0.70$, which agrees
well with the value $0.69$ reported in Ref.~\cite{Kiuchi:2014:41502} (albeit for 
their $70 \usk\meter$ resolution run). Also the disk mass of $0.06\usk M_\odot$ we 
found (at the same time) is identical to the value given in Ref.~\cite{Kiuchi:2014:41502}.
Therefore, the physical conditions for magnetic field amplification are very 
similar, apart from the different initial field strength.

In our run, the magnetic energy increases from ${\approx}10^{43}\usk\mathrm{erg}$ 
at merger time to ${\approx}10^{47}\usk\mathrm{erg}$ at the time of BH formation.
In Ref.~\cite{Kiuchi:2014:41502}, the energy is already at this level at merger time
and is amplified less than 1 order of magnitude in the $150 \usk\meter$ resolution run (in
stark contrast to their higher-resolution runs).
After collapse, the remaining energy outside the BH increases from 
${\approx}10^{47}\usk\mathrm{erg}$ up to almost ${\approx}10^{49}\usk\mathrm{erg}$,
at which point it saturates. In our simulation, the energy first stagnates around 
$5\times 10^{46}\usk\mathrm{erg}$, and then starts growing again around 
$30\usk\milli\second$ after merger, up to a value of $4\times 10^{49}\usk\mathrm{erg}$
reached $60\usk\milli\second$ after merger. 
We do not observe saturation at this amplitude, but we cannot rule it out at later times.
The reasons for the different behavior are unclear. Reference \cite{Kiuchi:2014:41502}
clearly demonstrated that a $150 \usk\meter$ resolution is insufficient
to resolve the field amplification in the disk, therefore the differences should not
be taken too seriously. That said, we notice that Ref.~\cite{Kiuchi:2014:41502} 
already reached a slightly higher magnetic energy directly after collapse, which makes 
it easier to resolve MRI effects in the disk. This might explain the delayed onset
of amplification in our case. 
For more details about our high-resolution run we refer to
Sec.~\ref{res}, where the differences with respect to our standard
resolution run are discussed.

An important statement in Ref.~\cite{Kiuchi:2014:41502} is that no coherent structure of 
the poloidal component was found. This contrasts with our results with a lower initial 
magnetic field. 
Comparing the field lines shown in Fig.~\ref{fig:field3d_final_H4_IF} to the ones in 
Fig.~1 of Ref.~\cite{Kiuchi:2014:41502}, we find indeed that the ``twister'' structure exposed 
in the former cannot be seen in the latter. 
On the other hand, the absence of a strongly collimated field along the BH axis reported in
Ref.~\cite{Kiuchi:2014:41502} agrees with our findings.
The apparent absence of the twister structure might also be an artifact of the different
selection of field lines and the larger scale of the plot in Ref.~\cite{Kiuchi:2014:41502}, resulting
in a lower field line density near the ``twister'' structure. Furthermore,
as described in the Appendix, we made an effort to avoid seeds in the less
regular regions between field lines of opposite direction.
For those reasons, and also because of the lower resolution of our run, the comparison
of the field structure remains rather inconclusive.
We note  however that our results do not rely solely on the field line plots. Using 
histograms in Fig.~\ref{fig:hist_final_H4_IF}, we demonstrate that the dependence of the
field energy on the $\theta$ coordinate is relatively flat and only falls off strongly 
between $50$--$30\degree$ around the spin axis.

Finally, we note the study \cite{Kiuchi:2015:1509.09205}, in which
additional refinement levels are added, down to a grid spacing of
$17.5 \usk\meter$, in order to resolve the Kelvin-Helmholtz (KH) instability during the
first few ms after merger. Those results show that a
much higher resolution than the one implemented in our simulations is
necessary in order to fully resolve the magnetic field amplification
due to the KH instability during merger. Therefore, the magnetic field
amplification inside the HMNS is most likely underestimated by our
runs. The question of how this influences the post-collapse phase is not
trivial, since an important fraction of the magnetic energy produced in the shear layer
is likely to be swallowed by the BH upon collapse.

\subsection{Unequal-Mass H4 model}
\label{sec:h4_q08}

For the H4 EOS, we found an enormous influence of the mass ratio on the
magnetic field amplification (see also Section \ref{res}).
The total magnetic energy and the maximum of the magnetic field are shown in 
Figures~\ref{fig:emag_eos_q} and~\ref{fig:Bmax_eos_q} in comparison to the
equal-mass H4 model as  well as the ideal-fluid models.
As one can see, the lifetime of the HMNS (${\approx}24\usk\milli\second$) for 
the H4 unequal-mass case is more than twice as long as that for the H4 equal-mass case.
During this phase, the field is growing exponentially, with the exception
of the last $5\usk\milli\second$ before collapse. The time scale of the 
exponential growth is also shorter than for the equal-mass case. 
Shortly before the collapse to a BH, the energy is around 4 orders of magnitude 
larger than for the equal-mass case, and the maximum field strength more than 2 orders 
of magnitude larger. The fact that these values do not change drastically during collapse
implies that most of the energy was contained in regions well outside the HMNS and that 
the field was also strongest there. 
As discussed in Section \ref{res}, we attribute at least part of this much stronger 
amplification to the magnetorotational instability.

The amplification after the collapse to a BH is comparable in growth rate to the ideal-fluid unequal-mass 
case (which showed a prompt collapse after merger). We conclude that the lifetime
of the HMNS is a very important factor for the post-collapse field strength
in the torus. It is likely that the large differences we see between the ideal-fluid and H4 unequal-mass 
cases are mostly due to the chosen total mass, i.e.~we expect more
similar results when comparing H4 and ideal-fluid EOS unequal-mass models with total
masses chosen such that the HMNS lifetime is the same. Parameters other than 
the HMNS lifetime, namely disk mass, BH spin, and accretion rate, 
are comparable to the \texttt{IF\_q08} case and cannot explain the much larger 
amplification.

The structure and distribution of the magnetic field $32 \usk\milli\second$
after merger is shown in Figures~\ref{fig:hist_final_H4_IF} 
and~\ref{fig:field3d_final_H4_IF}. Apart from the increased amplitude, we 
find that for the unequal-mass case, a larger fraction of the energy is contained
in the toroidal field near the equator. The field strength $B_{90}$ 
reaches $\sim 6 \times 10^{15}\usk\Gauss$ near the equator, more than 2 orders of magnitude
above the strength for the equal-mass case ($\sim 3 \times 10^{13}\usk\Gauss$).
The opening angle of the conical structure is also smaller.
As in the equal-mass case, the field near the axis does not contribute significantly
to the total magnetic energy, and the field strength $B_{90}$ near the axis is 
around 2 orders of magnitude below the equatorial value. Due to the overall increase
in amplitude however, this now corresponds to a field strength 
$B_{90}\approx 3\times 10^{13}\usk \Gauss$ near the axis.

\subsection{Influence of Resolution \label{res}}

 \begin{figure}
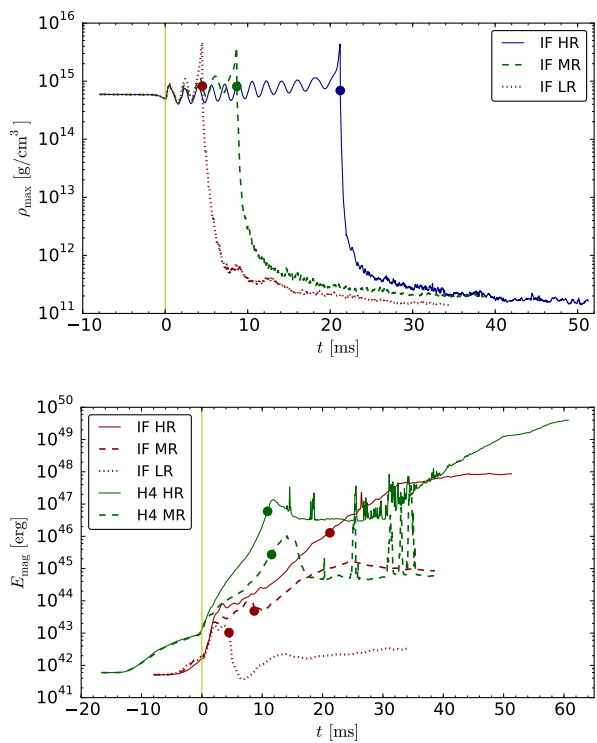

  \centering
  \includegraphics[width=0.95\columnwidth]{{{rhomax_conv_IF}}}
  \includegraphics[width=0.95\columnwidth]{{{emag_conv_all}}}
  \caption{Evolution of maximum rest-mass density (upper panel) and magnetic 
  energy (lower panel) for the equal-mass ideal-fluid model \texttt{IF\_q10\_UU} 
  at different resolutions, with finest grid spacing of $dx\approx177$, 222, 277 
  m for the high, medium and low resolutions, respectively. The evolution
  of the magnetic energy is also shown for the equal-mass H4 model \texttt{H4\_q10} with two 
  different resolutions: $dx\approx150$ m (HR) and 186 m (MR).}
  \label{fig:res}
\end{figure}

We performed simulations at different resolutions for the ideal-fluid and H4
equal-mass models (\texttt{IF\_q10\_UU} and \texttt{H4\_q10}).  First,
we discuss the ideal-fluid case, while the H4 case will be discussed
at the end of this section. In the last paragraph, given its
particular relevance, we will also discuss the impact of the chosen
resolution on the unequal-mass H4 model (\texttt{H4\_q08}).

Figure \ref{fig:res} shows the evolution of the maximum rest-mass density 
and magnetic energy at three different resolutions: $dx\approx 177$, 
$222$, and $277\usk\meter$ (where $dx$ is the finest grid spacing). 
The resolution affects the rest-mass density evolution only in the post-merger 
phase. The lifetime of the HMNS is extremely sensitive to small numerical
errors and numerical convergence is difficult to achieve. In our case, higher 
resolutions resulted in a longer lifetime, and we see no convergence for the 
employed resolution range. Note however that in general the lifetime of 
HMNSs also depends very strongly on their mass.
 
The HMNS lifetime directly influences the disk mass, because the strong 
oscillations of the HMNS in conjunction with the rapid rotation constantly
eject matter into the disk. Indeed, the disk mass increases from 
$0.015 \usk M_\odot$  for the lowest resolution (and shortest HMNS lifetime) 
to $0.077\usk M_\odot$ at the highest resolution. 
The mass and spin of the BH on the other hand are only weakly affected 
by the HMNS lifetime. The differences between high and medium resolution at
$30\usk\milli\second$ after collapse are both below $1.5\%$.

During the first ${\sim}2\usk\milli\second$ after merger, the magnetic energy
shown in the lower panel of Fig.~\ref{fig:res} exhibits an exponential increase,   
with a growth rate that depends only weakly on the resolution.
The saturation of this exponential growth on the other hand sets in later (and 
at higher energies) for higher resolution. 
This amplification is most likely associated (at least in part) 
with the KH instability, which can be captured only on scales larger 
than the grid spacing and therefore is not entirely accounted for 
in our simulations. 

In the subsequent evolution with medium and high resolution, the energy grows 
exponentially at comparable rate, but more slowly than directly after merger.
We can attribute this to amplification of the field in the disk, since the 
additional energy is obviously not swallowed into the BH during the collapse 
of the HMNS, and because the amplification continues after collapse until
it saturates. For the low resolution, the BH forms shortly after merger and 
the evolution of the field energy is due to the disk afterwards.
For all resolutions, the energy increase ceases at some point.
With increasing resolution, we observe a longer growth phase and a higher 
final amplitude. The difference between low and high resolution is more than 
5 orders of magnitude.
One possible explanation would be that the magnetic field amplification 
mechanism is acting also on small scales 
which are better resolved with a finer grid spacing.

\begin{figure}
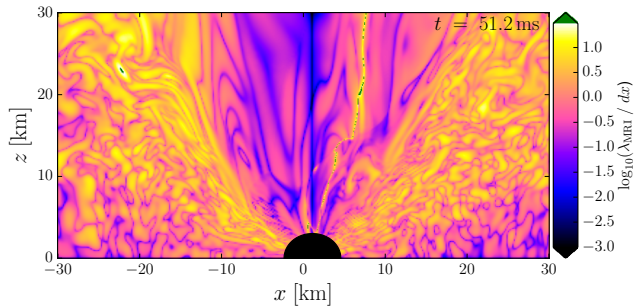

  \centering
  \includegraphics[width=1.01\columnwidth]{{{lambda_mri_snap_IF_30pts}}}
   \caption{Meridional view of $\lambda_\mathrm{MRI}/dx$ for the highest-resolution simulation ($dx\approx177$ m) of model \texttt{IF\_q10\_UU}, towards the end of the simulation ($t=51.2$ ms).}
  \label{fig:MRI}
\end{figure}

One such mechanism that could operate in the disk is the 
MRI. The wavelength of the fastest 
growing mode of the MRI is approximately given by
$\lambda_\mathrm{MRI} \approx (2\pi/\Omega) \times B_k/\sqrt{4\pi\rho}$, 
where $\Omega$ is the angular velocity and $B_k$ the magnetic field 
strength along the corresponding 
wave vector. In order to properly resolve this effect the finest grid spacing 
$dx$ has to cover $\lambda_\mathrm{MRI}$ with at least 10 points 
(see, e.g., Ref.~\cite{Siegel:2013:121302}). 
Figure~\ref{fig:MRI} shows the ratio $\lambda_\mathrm{MRI}/dx$ 
for the highest-resolution run ($dx \approx 177 \usk\meter$) 
at the end of the simulation. In this plot, the total magnetic field strength is 
used instead of $B_k$, and therefore the given ratio represents an upper limit. 
The ratio reaches maximum values ${\approx} 5$--$10$ along the conical structure 
separating the disk from the funnel, where the magnetic field is the strongest. 
This indicates that a resolution 
$dx\lesssim 100\usk\meter$ would be necessary in order to start resolving 
the MRI in that region. 
We note however that our formula for the wavelength does not take into account
general-relativistic corrections and uses an idealized disk model. 

Saturation of the amplification is not the only possible contribution
to the flattening of the magnetic energy growth that happens 
${\sim}15$--$20 \usk\milli\second$ after collapse.
Since the accretion time scale of the disk is ${\sim}50\usk\milli\second$, 
we can expect that the magnetic energy contained in the accreted matter is 
relevant. Assuming that the magnetic strength in the 
inner disk grows as fast as in the remaining disk, the net increase would be 
zero when the accretion time scale and growth time scale agree. On the other hand, the 
fact that the maximum field strength and $B_{90}$ saturate as well disfavors 
this scenario. Then again, the change of the disk structure due to accretion 
could affect the amplification mechanism, which would make the outcome
sensitive again to the time of the collapse. The picture is complicated even more
by the differences in disk mass due to the different HMNS lifetimes. 
For those reasons we cannot conclusively 
associate the flattening of the magnetic energy evolution to an actual 
saturation of the involved magnetic field amplification mechanisms.

The final magnetic energy between medium and high resolution differs
by about 3 orders of magnitude, with the highest-resolution case
reaching an increase of more than 6 orders of magnitude in
$E_\mathrm{mag}$ compared to the beginning of the simulation. This
amplification factor should be regarded as a lower limit that might be
overcome with even higher resolution.

We now turn our attention to the H4 equal-mass model. In this case, 
we performed simulations at two different resolutions $dx\approx186\usk\meter$ 
(MR) and $150 \usk\meter$ (HR). The latter corresponds to the grid spacing employed 
in the lowest-resolution run of Ref.~\cite{Kiuchi:2014:41502} for a very similar 
model. A direct comparison has already been presented in Section \ref{sec:kiuchi}.
The lower panel of Fig.~\ref{fig:res} shows the evolution of the magnetic 
energy for the two H4 simulations. In contrast with the ideal-fluid case, there is 
no significant difference in the time of collapse to a BH (circle markers). 
Prior to collapse, the magnetic field amplification is stronger in the higher-resolution 
case, indicating that the dominant amplification mechanisms are 
not fully resolved. As for the ideal-fluid case, we estimated $\lambda_\mathrm{MRI}/dx$
and found that only some isolated lumps inside the ``twister'' structure
are resolved with more than 10 grid points for the high-resolution case.
In the highest-resolution run, a further increase in magnetic energy 
is observed some time after BH formation, corresponding to a strong amplification 
in the accretion disk. The simulation 
stops about $50 \usk\milli\second$ after collapse and we find an overall change in magnetic 
energy of almost 8 orders of magnitude compared to initial data. This corresponds 
to an average increase of the magnetic field strength of about $4$ orders of 
magnitude and it could be even larger with higher resolution.
 
\begin{figure}
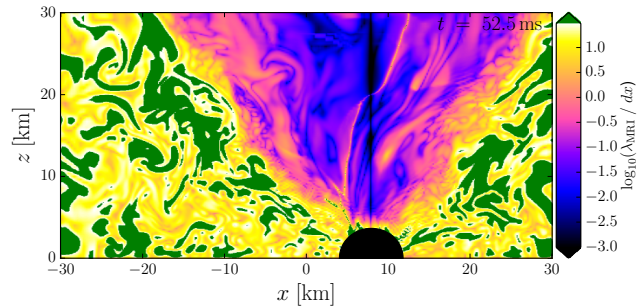

  \centering
  \includegraphics[width=1.01\columnwidth]{{{lambda_mri_snap_H4_q08_t30_30pts}}}
   \caption{Same as Figure~\ref{fig:MRI} for model \texttt{H4\_q08} at resolution $dx\approx186$ m and at $t=52.5$ ms.}
  \label{fig:MRI_H4q08}
\end{figure}

For the unequal-mass H4 model we performed only one simulation with a
finest grid spacing of $dx\approx186$ m. Nevertheless, because the model
shows by far the strongest magnetic field amplification
(c.f.~Fig.~\ref{fig:emag_eos_q}), it is important to assess how well
the MRI is resolved in this case.  As shown in
Figure~\ref{fig:MRI_H4q08} and differently from all other models in
this study, at the end of the simulation $\lambda_\mathrm{MRI}/dx>10$
almost everywhere in the accretion disk. We attribute this to the fact
that the magnetic field strength becomes higher because of the much
longer lifetime of the HMNS and this makes $\lambda_\mathrm{MRI}$
larger.  In turn, the MRI is better resolved, leading to a stronger
amplification and thus to an even stronger magnetic field. This
positive-feedback process provides a likely explanation for the fact that
this particular model ends up with a magnetic energy that is several
orders of magnitude higher.  However, future simulations at higher
resolution will be necessary in order to confirm this picture.

\begin{figure*}
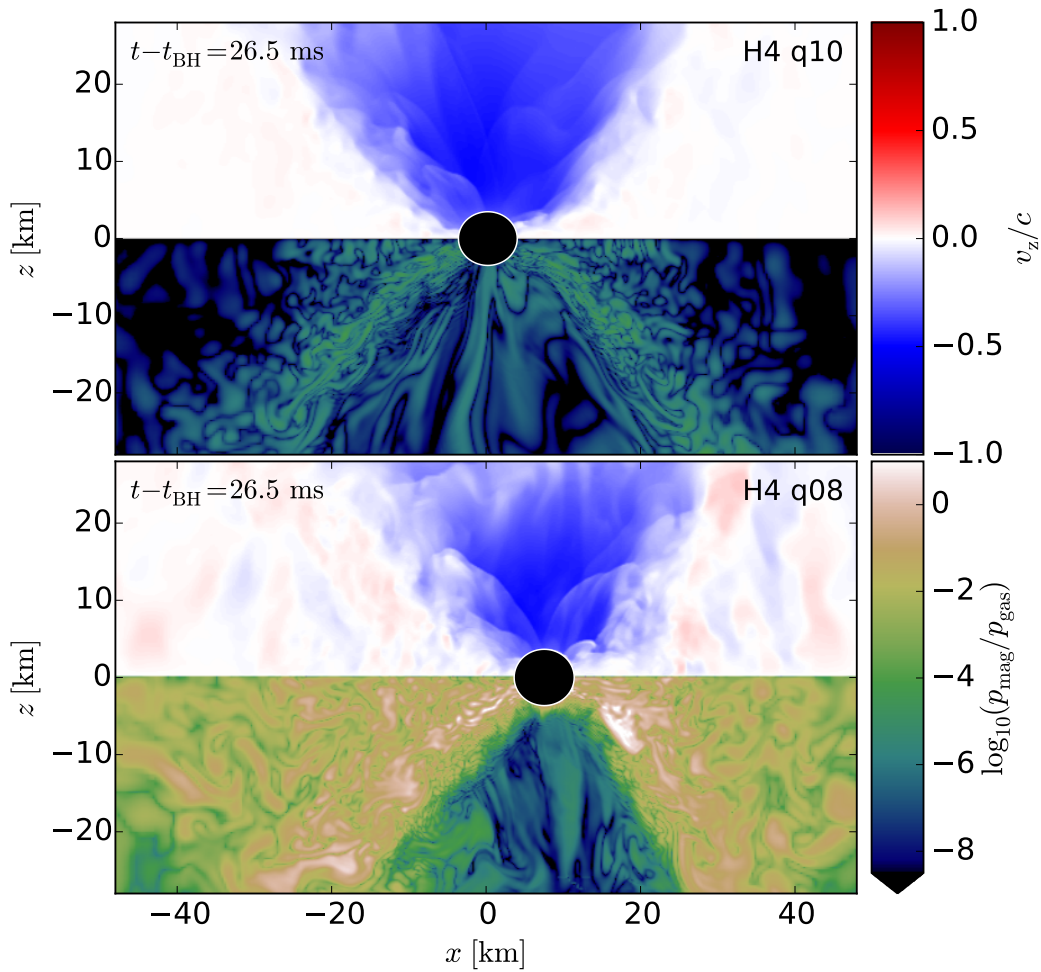

  \centering
  \includegraphics[width=0.80\textwidth]{{{velz-beta-H4}}}
  \caption{Meridional view of BH and accretion torus for the equal- and unequal-mass H4 simulations. The panels refer to 26.5 ms after BH formation and show in the top half ($z>0$) the fluid velocity along the $z$ axis and in the bottom half ($z<0$) the magnetic-to-fluid pressure ratio in log scale.}
  \label{fig:velz-beta-H4}
\end{figure*}

\section{Short Gamma-Ray Bursts and Other Electromagnetic Signals \label{sec_SGRB}}

The possibility that the merger of two NSs may be
accompanied by an SGRB has been discussed for
several decades (see, i.e.,
Refs.~\cite{Paczynski1986,Eichler:1989:126,Narayan1992,Ruffert:1999:573}). 
The generally invoked scenario is
one in which the merger product is a BH surrounded by a
massive accreting torus. The rapid accretion of the disk onto the
newly formed BH provides the central engine for the burst.  Another
possibility that has been suggested for powering the engine is the
electromagnetic spindown emission from a highly magnetized NS
(see, i.e., Refs.~\cite{Dai2006,Metzger2008}), which survives for
some time before collapsing to a BH or remains as a stable NS (if
allowed by its mass \cite{Giacomazzo2013ApJ...771L..26G}). Finally, an
alternative ``time-reversal'' scenario has been proposed 
\cite{Ciolfi:2015:36,Ciolfi:2015:PoS} in which the NS survives for a long time (up to
spindown time scales) before eventually collapsing to a BH, and while
its rotational energy powers a long-lasting X-ray signal (potentially
explaining the X-ray afterglows commonly observed by {\it Swift}; see,
e.g., Ref.~\cite{Rowlinson2013}), the SGRB itself is powered by accretion
onto the resulting BH, as in the standard scenario. In this work we
focus on the first, most studied case in which a BH is formed in less
than $100 \usk\milli\second$ after merger.

The $\gamma$-ray emission is believed to be produced within a
relativistic outflow (at the distances at which this becomes
optically thin), and hence a crucial ingredient of any SGRB model is
its ability to drive a jet. Two main mechanisms have been invoked:
neutrinos (see, e.g., Ref.~\cite{Ruffert:1999:573}) and magnetic fields.  At high
accretion rates, neutrinos can, in principle, tap the thermal energy
of the disk produced by viscous dissipation and liberate large amounts
of its binding energy via the $\nu\bar{\nu}\rightarrow e^+ e^-$ process
in regions of low baryon density. However, recent simulations 
of the hyperaccreting disk that include neutrino transfer
have shown that, if the remnant torus and environment is that of a
BNS merger, then neutrino emission is too short and too weak to
yield enough energy for the outflow to break out from the surrounding
ejecta as a highly relativistic jet \cite{Just2016}. Hence, it has been 
concluded that neutrino annihilation alone cannot power SGRBs from
BNS mergers.

On the other hand, a strong poloidal magnetic field
around a spinning BH can extract rotational energy and power
an outflow \cite{Blandford1977}. This mechanism is commonly 
considered the most viable one for producing jets. Therefore, the topology of
the post-merger magnetic field in our simulations plays an especially
important role. Evidence for a geometrical structure compatible with 
jet formation in the merger of a BNS was found in Ref.~\cite{Rezzolla:2011:6}, 
although as already discussed earlier only recently it was possible to show that BNS mergers can 
actually produce an ``incipient jet'' along the spin axis of the resulting BH, 
defined as a collimated and mildly relativistic outflow that is at least 
partially magnetically dominated \cite{Ruiz2016}. A similar 
result was obtained earlier for NS-BH binary mergers 
\cite{Paschalidis:2015:14}.

Our simulations show the formation of a spinning BH with
spin parameter in the range ${\sim}0.6$--$0.8$ (see Table~\ref{tab:final-system}) and 
surrounded by a torus of at least a few percent of
a solar mass, with the unequal-mass models yielding the larger torus masses.
These results are consistent with previous results 
(e.g., Refs.~\cite{Rezzolla:2010:114105,Rezzolla:2011:6}). 
The average accretion rates are of the order of ${\sim}1\usk M_\odot\usk\second^{-1}$. 
For typical conversion efficiencies of accreted mass to observed radiation, 
these accretion rates and torus masses satisfy the energy requirements of 
the observed SGRBs, in particular in the unequal-mass cases 
\cite{Giacomazzo2013}.
However, the ability to launch a magnetically driven jet requires, in addition 
to a massive disk, also a strong poloidal field along the spin axis of the BH. 

As discussed in the previous sections, in our simulations magnetic fields 
are strongly amplified after merger during the HMNS lifetime (see Figs. 
\ref{fig:emag_eos_q}, \ref{fig:Bmax_eos_q}, and \ref{fig:Bmean_eos_q}). Magnetic field amplification 
continues in the disk after BH formation although in some cases an overall decrease 
of magnetic energy is observed, possibly due to accretion.
As a result of the amplification, and in particular of the winding of the magnetic field 
lines, the toroidal component becomes dominant over the poloidal one in the disk.
Along the edge of the accretion torus we observe the development of a mixed 
poloidal-toroidal ``twister'' structure.
For the unequal-mass H4 model, we observed a particularly strong amplification
of both the poloidal and toroidal components.
For this case, the density-weighted mean value grows by over 2 orders
of magnitude (see Fig.~\ref{fig:Bmean_3mag_eos_q}).
One important reason for this difference lies in the
fact that, for this combination of EOS and NS masses, the
HMNS formed upon merger survives for a much longer
timescale compared with the other cases that we studied (see Sections 
\ref{sec:h4_q08} and \ref{res}).
The higher torus mass and the stronger magnetic field amplification 
make the H4 unequal-mass case the most favorable of our models to produce a jet. 
Also the magnetic field morphology and the half-opening angle of the funnel 
(smaller than $30\degree$) are compatible with what is needed to
drive a SGRB (see Fig.~\ref{fig:field3d_final_H4_IF} and \ref{fig:2d_rms_cut_xz_H4}).

Figure~\ref{fig:velz-beta-H4} shows the fluid velocity along the orbital axis and the magnetic-to-fluid 
pressure ratio\footnote{The ratio is defined as $\beta\equiv b^2/(2p)$, where 
$b^2\equiv b^\mu b_\mu$  and $b^\mu$ is the 4-vector of the magnetic field 
as measured by the comoving observer \cite{Giacomazzo:2007:235}.}
for the equal- and unequal-mass H4 simulations, $26.5 \usk\milli\second$ after BH formation.
In both cases matter 
inside the funnel and along the spin axis of the BH is still infalling and 
in the unequal-mass case the pressure ratio indicates that the fluid is becoming 
magnetically dominated at the 
edges of the disk, but inside the funnel magnetic field pressure is subdominant. 
In conclusion, despite the fact that some favourable conditions are met, we do not find evidence 
of jet formation. 
Our results confirm the expectation that unequal-mass 
systems produce more massive disks (for the same total baryonic mass) and 
we find that longer-lived HMNSs can lead to a much stronger magnetic field amplification, 
which might also support the formation of a jet.

From our results, we are not in a position to exclude that the systems under investigation 
can form a jet.
Our present simulations are limited to less than $30 \usk\milli\second$ (in one case $50$~ms) 
after BH formation and an 
outflow might still emerge on longer time scales. Moreover, magnetic field amplification 
mechanisms that act 
on scales that are too small to be properly resolved with our present resolution (such as the Kelvin-Helmholtz instability) would 
provide much stronger amplification 
(see, e.g., Refs.~\cite{Kiuchi:2015:1509.09205,Kiuchi:2014:41502,Siegel:2013:121302}) 
and thus influence the dynamics.

Our simulations lack a neutrino treatment. As such, we
cannot compute the contribution of neutrinos to cooling and heating of the
remnant disk. Most importantly, our simulations do not allow us to
investigate the emergence of a jet driven by neutrino annihilation.
However, as discussed above, Ref.~\cite{Just2016} concluded that 
for the BNS merger scenario to yield a SGRB, jets must be magnetically
driven. 
Lacking neutrinos in our treatment should not 
prevent the simulations from showing the emergence of such a magnetic jet.
Nevertheless, neutrinos can still have an impact on the evolution of 
both the HMNS and the accretion disk. 

\begin{figure*}[!ht]
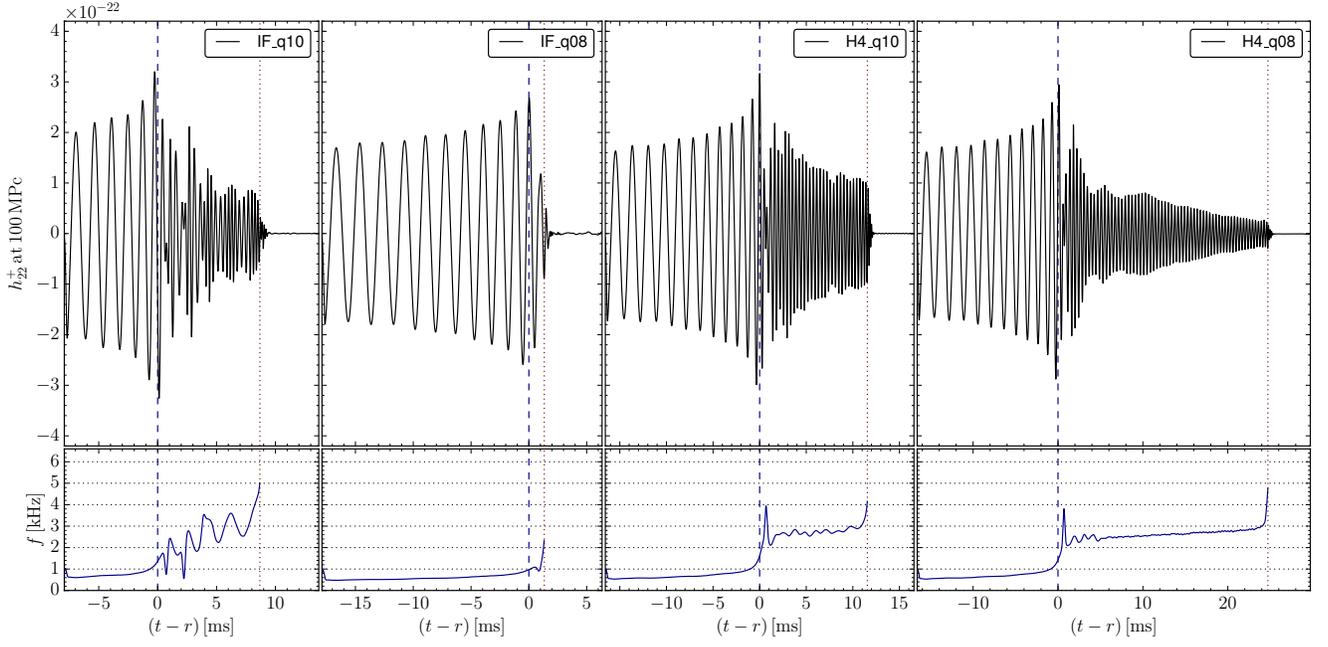

  \centering
  \includegraphics[width=0.99\linewidth]{{{gw_strain_multisim}}}
  \caption{GW signal for models (from left to right) \texttt{IF\_q10}, \texttt{IF\_q08}, 
  \texttt{H4\_q10} and \texttt{H4\_q08}.
  The top panels show the strain at nominal distance of $100$ Mpc.
  The lower panels show the instantaneous frequency.} 
  \label{fig:gw_strain_eoscomp}
\end{figure*}

\begin{figure*}[!ht]
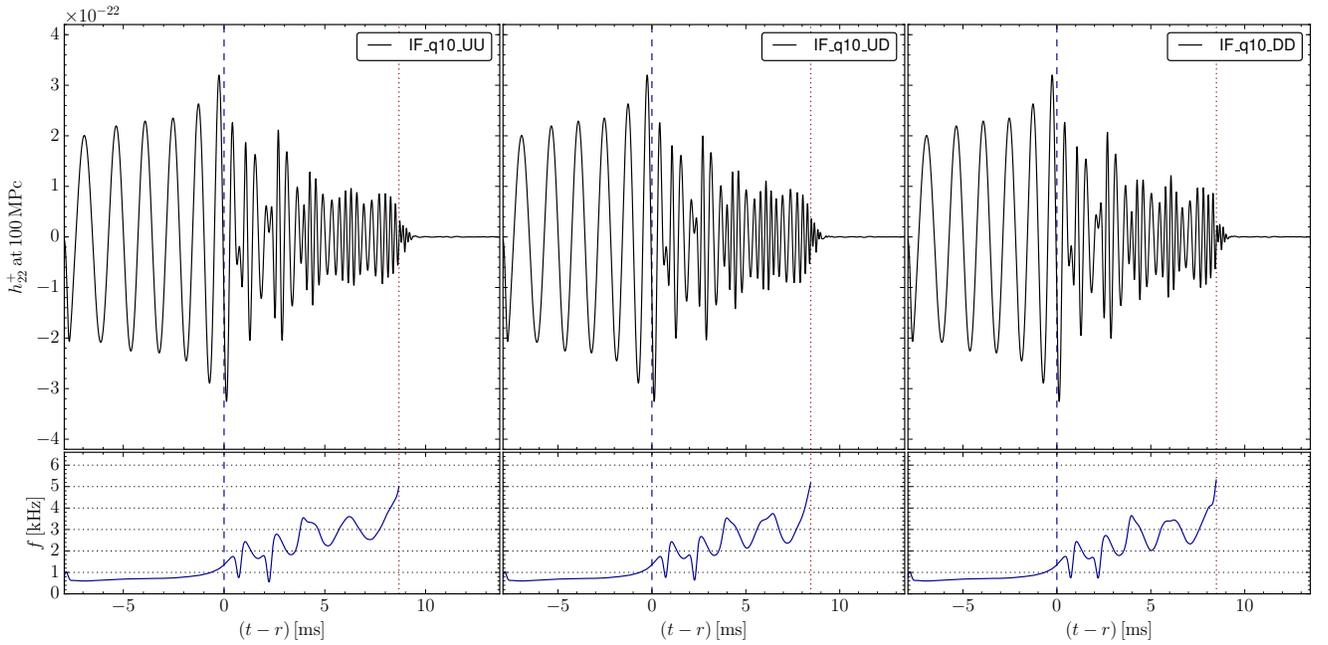

  \centering
  \includegraphics[width=0.99\linewidth]{{{gw_strain_UU_UD_DD}}}
  \caption{GW signal for models (from left to right)
    \texttt{IF\_q10\_UU}, \texttt{IF\_q10\_UD} and
    \texttt{IF\_q10\_DD}. The top panels show the strain at nominal
    distance of $100$ Mpc. The lower panels show the instantaneous
    frequency.}
  \label{fig:gw_strain_UU_UD_DD}
\end{figure*}

In addition to the prompt $\gamma$-ray emission produced within the
relativistic outflow and the associated X-ray and optical afterglows, the merger of two
NSs is also expected to create a significant amount of neutron-rich
radioactive elements, whose decay should result in a transient signal,
the so-called ``kilonova" or ``macronova'', in the days following the burst
(see, e.g., Refs.~\cite{Rosswog2005,Metzger2010}).  The emerging radiation is
expected to peak in the near IR, due to the large optical opacity of
the heavy $r$-process elements, and to be nearly isotropic.  As
such, it constitutes an interesting complement to the prompt gamma-ray
emission, which is expected to be generally beamed. 
Kilonova candidates were found to be associated with GRB~130603B
\cite{Tanvir2013}, a SGRB at redshift $z=0.356$, with GRB~060614~\cite{Yang2015, Jin2015}, and with GRB~050709~\cite{Jin2016}.
Another promising electromagnetic signal from BNS mergers is the isotropic X-ray 
emission powered by the spindown of a long-lived NS remnant 
\cite{Siegel:2016a,Siegel:2016b}, 
although such a signal is not expected if a BH is formed shortly ($<1$~s) after merger.

The observation of SGRBs or other electromagnetic counterparts in combination with 
the BNS merger GW signal will dramatically improve the scientific 
output of a detection. In the following section we discuss the GW emission from the 
BNS mergers studied in this work.

\section{Gravitational Waves \label{sec_GWs}}

For all runs we extract the GW signal at a fixed radius of $\sim 1100\,
\mathrm{km}$ via the Moncrief formalism (signal is extracted also via
the Weyl scalar $\Psi_4$, but only for cross-checking purposes). Note
that extrapolation at infinity is not performed for any of our
simulations.

In this section we present the strain of the GW signal as $h_{lm} =
h_{lm}^+ + ih_{lm}^\times$, namely, the coefficients of the spin-weighted
spherical harmonics expansion. 
In order to obtain the actual strain that would be measured by a GW
detector, one should multiply our value by the spin-weighted
spherical harmonics in order to take into account the signal
direction.
For each simulation we also extracted the instantaneous
frequency of the GW from the phase velocity of the complex strain, which is
shown in the bottom panels of Figs.~\ref{fig:gw_strain_eoscomp} and \ref{fig:gw_strain_UU_UD_DD}.

In Fig.~\ref{fig:gw_strain_eoscomp} we show the $l=m=2$ component of
the GW strain for models \texttt{IF\_q10}, \texttt{IF\_q08},
\texttt{H4\_q10} and \texttt{H4\_q08}. While in the \texttt{IF\_q08}
case, where the system promptly collapses to a BH, the GW includes
only inspiral, merger, and ringdown, in all the other cases a HMNS is
formed and therefore we have also a longer post-merger GW signal. In
the \texttt{IF\_q10} case the GW frequency during the HMNS lifetime
varies continuously. This behavior differs from the \texttt{H4} cases,
where the HMNS phases show signals with a very strong peak at specific
frequencies. Note that in the \texttt{H4} cases the HMNS has a longer lifetime, and
in the \texttt{H4\_q10} case the post-merger GW signal also has a
stronger amplitude with respect to the other models. As previously discussed,
however, the lifetime of the remnant also depends on the
resolution, with the HMNS surviving longer with higher resolution.

In terms of frequency, the H4 models show a drift towards higher
frequencies during the post-merger phase, which is more evident in the
\texttt{H4\_q08} case, where the remnant lasts longer and the value of
the frequency oscillates less. In Table~\ref{tab:final-system} we
report for all models the frequency at merger $f_\mathrm{merger}$ and,
for the H4 cases, also $f_\mathrm{HMNS}$, which indicates the
frequency corresponding to the most prominent post-merger peak in the
GW spectrum (called $f_\mathrm{peak}$ in Ref.~\cite{Bauswein:2012:11101} and 
$f_2$ in Ref.~\cite{Takami:2014:91104}). 
We do not provide $f_\mathrm{HMNS}$ for the ideal-fluid
models since \texttt{IF\_q08} has no HMNS remnant (it promptly
collapses to a BH) and in \texttt{IF\_q10} the frequency oscillates
too much to get an accurate estimate, as it is shown from both
the amplitude and spectral behaviors.

We also studied whether the effect of magnetic field orientation had
any impact on the GW signal. As shown in Fig.~\ref{fig:gw_strain_UU_UD_DD}, 
this impact is minimal. This may change if the magnetic field
is amplified to much larger values during merger.

\begin{figure}[!ht]
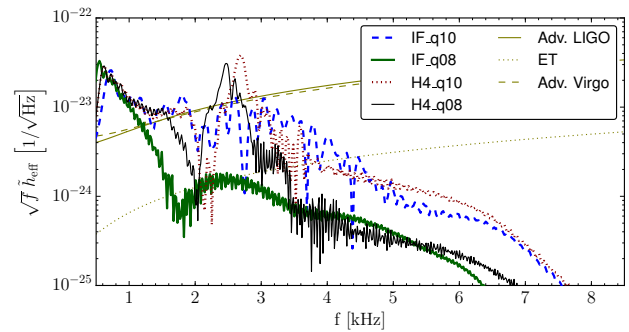

  \centering
  \includegraphics[width=0.99\columnwidth]{{{gw_spec_multisim_l2m2}}}
  \caption{GW spectra (solid lines) for the four models of Fig.~\ref{fig:gw_strain_eoscomp}
   in comparison to the sensitivity curves of GW detectors (dashed lines). The strain is given at distance 
   of $100$ Mpc.} 
  \label{fig:gw_spec_eoscomp}
\end{figure}

\begin{figure}[!ht]
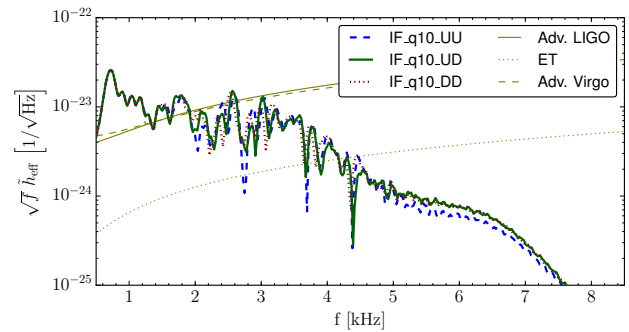

  \centering
  \includegraphics[width=0.99\columnwidth]{{{gw_spec_UU_UD_DD_l2m2}}}
  \caption{GW spectra (solid lines) for the three models of Fig.~
  \ref{fig:gw_strain_UU_UD_DD} in comparison to the sensitivity curves of GW detectors (dashed lines).
  The strain is given at distance of $100$ Mpc.} 
  \label{fig:gw_spec_UU_UD_DD}
\end{figure}

Finally, in Figures~\ref{fig:gw_spec_eoscomp} and \ref{fig:gw_spec_UU_UD_DD}
we plot the power spectra of the GW signals for all our
simulations against present and future ground-based detector
sensitivities (namely Advanced Virgo, Advanced LIGO, and the Einstein
Telescope, all in the standard broadband configuration).

The power spectrum we show in the plots is given by $h_{eff} (f) =
\sqrt{\tilde{h}^2_+(f) + \tilde{h}^2_\times (f)}$, where $\tilde{h}_+$
and $\tilde{h}_\times$ are the Fourier transforms of $h_{lm}^+$ and
$h_{lm}^\times$ for $l=m=2$. From both Figures we can see that the
inspiral phase would be detected by both Advanced Virgo and Advanced
LIGO for all models. Moreover, in Fig.~\ref{fig:gw_spec_eoscomp} we see
that for the H4 models the post-merger peak of the signal due to HMNS
oscillations would also be strong enough to be detected by Advanced LIGO
and Virgo.  If detected, this peak could play a very 
important role in constraining the NS EOS \cite{Bauswein:2012:11101,
  Takami:2014:91104, Takami2015}.

\section{Conclusions \label{sec_conclusions}}

In this paper we started our investigation of the magnetic field
structure formed in the post-merger of high-mass BNS systems, i.e., of
systems that produce a BH on a dynamical time scale after merger. We
focused in particular on two different EOSs, ideal fluid and H4, both
of which were used recently by other groups to study the merger of
equal-mass systems~\cite{Rezzolla:2011:6, Kiuchi:2014:41502,
Ruiz2016}. We have extended those previous investigations by including
also unequal-mass BNSs and by changing, for one configuration, also
the initial magnetic field orientation.

Compared to previous work, here we have introduced a more
systematic way to study the magnetic field structure in order to
better understand whether an ordered poloidal field is formed after
the merger or not. This has important consequences on the possible
formation of relativistic jets and on the central engine of SGRBs.

The main result of this work is that we observe the formation of an
organized magnetic field structure after the formation of a BH
surrounded by an accretion disk. This happens independently of EOS,
mass ratio, and initial magnetic field orientation. 
The main difference with what was reported by Rezzolla 
et al.~\cite{Rezzolla:2011:6} is that the field along the BH axis is 
neither strong nor strongly collimated. We observe a strong field near 
the edge of the torus, which
is not composed of
straight magnetic field lines, but instead has a more helical structure,
similar to the one observed in Ref.~\cite{Ruiz2016}. The initial magnetic
field orientation does not produce large differences, but we point out
that the \texttt{UD} configuration is the one leading to the smallest amount of
magnetic energy and the smallest values for $B_{90}$ along the conical 
structure separating the low-density funnel and 
the higher-density disk, where the magnetic field amplification is generically 
found to be the most efficient. The largest magnetic 
field is obtained in the unequal-mass model evolved with the H4 EOS
({\tt H4\_q08}). This is due to the much longer HMNS phase in this
case which allows for a much larger magnetic field amplification 
(likely contributed by a better-resolved MRI, c.f.~Sec.~\ref{res}).

We did not observe the formation of a jet in any of the simulations, 
consistently with what was
seen in Refs.~\cite{Rezzolla:2011:6, Kiuchi:2014:41502}, but this is not
unexpected considering the recent results of Ref.~\cite{Ruiz2016}. It is
indeed known that a magnetically dominated region in the BH ergosphere
is a necessary condition for the activation of the BZ
mechanism~\cite{Blandford1977}. On the one hand, our resolution is in general 
not high enough to be able to fully resolve the KH instability during merger
and the MRI after merger (with the possible exception of model {\tt H4\_q08}, 
see Sec.~\ref{res}), and therefore the magnetic 
field amplification might not be strong enough to activate the BZ mechanism. 
On the other hand, our simulations are limited to a few tens of ms after BH 
formation, while it could take longer to realize the conditions to form a jet
\cite{Ruiz2016}.

A recent study~\cite{Parfrey2015} investigated a mechanism where magnetic loops
drifting into the BH are inflated and forced to open due to differential rotation
between disk and BH, potentially powering jets. The study assumed the force-free MHD 
limit as well as axisymmetry, and required a critical size of the initial loops for the 
case of prograde disks. Therefore it is not clear if this mechanism plays a role in our 
setup. Future studies can help in assessing the viability of this scenario.

Our next step will be to employ the analysis techniques
developed in this paper to study the same (or similar) systems when 
evolved with our subgrid model~\cite{Giacomazzo:2015} or with 
resolutions that are high enough to better capture KH and MRI. 
Moreover, we will evolve for a longer time after BH formation.
Since in this paper we have shown that the magnetic field
structure is qualitatively the same independently of the EOS, mass ratio,
and magnetic field orientation, we expect the results
of Ref.~\cite{Ruiz2016} to be general and we will assess this statement in
future simulations.

Another important ingredient will be the use of finite-temperature
EOSs and neutrino emission, which were included only recently in GRMHD
simulations by another group~\cite{Palenzuela2015}. These will not
produce qualitatively different results, but they will provide a more
accurate description of the post-merger phase and GW emission. 

Our step-by-step study will help in assessing the individual
contributions of the different physical ingredients (high magnetic
fields, finite-temperature EOSs, and neutrino emission) to the possible
emission of relativistic jets and SGRBs.

Initial data used for the simulations described in
this paper, as well as gravitational wave signals and movies from our
simulations are publicly available online.

\begin{acknowledgments}
We acknowledge support from MIUR FIR grant No.~RBFR13QJYF. We also
acknowledge PRACE for awarding us access to SuperMUC based in Germany
at LRZ (grant GRSimStar). Numerical simulations were also run on the
clusters Fermi and Galileo at CINECA (Bologna, Italy) via INFN
teongrav allocation and via ISCRA grants IsC34\_HMBNS and
IsB11\_MagBNS. T.K. and B.G. acknowledge partial support from
``NewCompStar", COST Action MP1304. R.P. acknowledges support from
award NSF-AST 1616157. L.B. acknowledges partial support from JSPS
Grant-in-Aid for Scientific Research(C) No. 26400274.
\end{acknowledgments}

\appendix

\section{Visualizing the Field Structure}
\label{sec:viz_field}

Our visualization method for the magnetic field aims at solving the following
problems. First, the magnetic field in our simulations is organized in 
tubes, and the direction of the field between neighbouring tubes changes sign.
The in-between field is typically weaker and less regular. Using random or 
regularly spaced seed points for the integration is bound to miss the 
strong field regions. Second, showing the field lines everywhere leads to visual 
clutter and obscures the global structure. We therefore have to choose a smaller 
number of field lines which are representative of the structure. It is important 
to use a well-defined automated method for the fiel dline filtering since a 
biased selection can result in misleading plots. Finding a good selection rule is
difficult because the field strength varies strongly between the different 
parts of the field we are interested in.

To solve the first problem, we divide the volume of interest into a coarse 
grid ($15^3$ cells). In each cell, we determine the 
location of maximum field strength and use it as a seed point. 
We then integrate the field lines for all seed points.
The solution of the second problem is more involved. First, we divide our domain
into bins regularly spaced in $\cos(\theta)$, where $\theta$ is the angle 
between the BH axis and the position vector relative to the BH.
We then sort the field lines in each bin by their maximum field strength inside
the given bin. Next, we assign to each field line the maximum of its rank in 
all bins it traverses. We then sort the field lines by this ``maximum local 
importance'' measure and keep only a given number of them.

This prescription results in a balanced distribution of field lines in 
the different parts of the field (axis, disk, torus) despite a strongly 
varying strength on both large and small length scales.
One could argue however that the binning in terms of $\cos(\theta)$ might
highlight conical structures where there are none in reality. For example,
the strong field in the torus casts a ``shadow'' radially outwards
where weaker field lines are not shown. To validate that the visual 
impression given by the 3D plots shown in this article is correct, we also
compared different visualizations, such as volume rendering of the field 
strength and simple 2D cuts.

\bibliographystyle{apsrev4-1}
\bibliography{trento}

\end{document}